\documentclass{aa}  

\usepackage{graphicx}
\usepackage{amsmath}
\usepackage{amsfonts}
\usepackage{amssymb}
\usepackage{times}
\usepackage{txfonts}
\usepackage{lipsum}
\usepackage{subcaption}         
\usepackage{lscape}             
\usepackage{placeins}           

\usepackage{upgreek}
\usepackage{multicol}
\usepackage{mathrsfs}
\usepackage{pifont}
\usepackage{amscd}
\usepackage{latexsym}
\usepackage{geometry}
\usepackage{color}
\usepackage{fancyhdr}

\usepackage{multirow} 
\usepackage{float}
\usepackage{chngcntr} 
\usepackage{enumitem}
\usepackage{xcolor}
\usepackage{hyperref}
\usepackage{graphicx}   
\usepackage{dcolumn}    
\usepackage{bm}         

\newcommand{\msm}{ \mathcal{M} {\rm\tiny sm} }

\begin{document}


   \title{The Shear-to-Cosmology Paradigm}
   \subtitle{I. Hybrid Field-Level and Simulation-Based Framework \\ for Weak Lensing Surveys}

%

   \author{Jiacheng Ding\inst{1}
        \and Chen Su\inst{1}
        \and Ji Yao\inst{1}\fnmsep\thanks{E-mail: ji.yao@shao.ac.cn}
        \and Le Zhang\inst{2,3}\fnmsep\thanks{E-mail: zhangle7@mail.sysu.edu.cn}
        \and Huanyuan Shan\inst{1}\fnmsep\thanks{E-mail: hyshan@shao.ac.cn}
        }

   \institute{
   Shanghai Astronomical Observatory, Chinese Academy of Sciences, Shanghai 200030, P. R. China
   \and 
   School of Physics and Astronomy, Sun Yat-Sen University, Zhuhai 519082, P. R. China
   \and
   CSST Science Center for the Guangdong–Hong Kong–Macau Greater Bay Area, SYSU, Zhuhai 519082, P. R. China
   }

   \date{Received September 30, 20XX}


    \abstract
    {Precise cosmological inference from next-generation weak-lensing surveys requires non-Gaussian information beyond two-point statistics, while conventional field-level analyses based on convergence reconstruction can introduce additional systematics.}
    {This work aims to develop a scalable and robust framework that directly maps shear fields to cosmological parameters, exploiting non-Gaussian information while avoiding convergence reconstruction to improve constraints from Stage-IV weak-lensing surveys.} 
    {We propose a hybrid machine-learning (ML) framework that combines field-level inference (FLI) with simulation-based inference (SBI). The FLI network performs field-level compression by extracting informative, non-Gaussian features directly from shear field, which are then passed to SBI to model complex posteriors. To mitigate noise from intrinsic galaxy shapes, we further develop a blind, training-free, PCA-based shear denoising method.}
    {Using CSST-like mock catalogs, we show that shear-based inference achieves an approximately twofold improvement in FoM over convergence-based approaches. Furthermore, within the shear-based framework, the combination of PCA denoising and ML compression yields a 36.4\% FoM improvement relative to standard two-point statistics.}
    {This work establishes a scalable framework for shear-based cosmological inference that unlocks field-level non-Gaussian information and significantly improves constraints for Stage-IV weak-lensing surveys.}
    
    
    \maketitle
    \nolinenumbers 

\section{Introduction}
\label{sec:introduction}

Weak lensing (WL) is a powerful probe in modern cosmology. According to General Relativity, light rays are deflected when passing through inhomogeneous matter distributions, leading to subtle distortions in the observed shapes of background galaxies. By statistically analyzing the shape distortions of millions of galaxies, one can directly map the large-scale distribution of total matter, including both dark matter and baryons, across the cosmic web. This makes WL an essential tool for constraining cosmological parameters, testing theories of gravity, and investigating the nature of dark energy~\citep{2015RPPh...78h6901K, 2018ARA&A..56..393M, 2022iglp.book.....M}. In addition, WL-based dark matter (DM) reconstruction techniques~\citep{2007Natur.445..286M, 2013MNRAS.433.3373V, 2014MNRAS.442.2534S, 2018PASJ...70S..26O} and DM halo identification methods~\citep{2004ApJ...606...67H, 2005ApJ...635...60T, 2012ApJ...748...56S} have played a key role in tracing the growth and spatial distribution of large-scale structures (LSS) in the Universe.
\par
Over the past decade, Stage-III WL surveys such as DES~\citep{2016MNRAS.460.1270D}, KiDS, and HSC~\citep{2018PASJ...70S...4A} have mapped thousands of square degrees of the sky with dedicated pipelines and achieved $1\%$ control of systematics, establishing a solid foundation for precision cosmology. Building on these advances, ongoing and upcoming Stage-IV surveys, including the China Space Station Survey Telescope (CSST)~\citep{2024MNRAS.527.5206Y,2025arXiv250704618C,2025SCPMA..6880402G}, Euclid~\citep{2025A&A...697A...1E}, LSST~\citep{2019ApJ...873..111I}, and the Roman Space Telescope~\citep{2015arXiv150303757S}, will expand sky coverage to tens of thousands of square degrees and deliver substantially improved measurement precision, enabling unprecedented tests of dark energy, gravity, and structure formation.
\par
Traditionally, cosmological inference has relied on two-point statistics within a Bayesian framework~\citep{2020A&A...641A...6P, 2021MNRAS.506L..85L, 2021Univ....7..213P, 2022PhRvD.106f3536T}. However, several challenges become increasingly severe as data quality improves. Achieving Stage-IV accuracy in modeling WL observables is particularly demanding due to nonlinear gravitational evolution, baryonic effects, and complex observational systematics, including PSF modeling, survey geometry, photometric redshift uncertainties, and intrinsic alignments~\citep{1999ApJ...522L..21H, 2002A&A...396....1S, 2009ApJ...695..652B, 2010A&A...516A..63S, 2013MNRAS.431.1547B,2023A&A...673A.111Y}.
\par
In addition, late-time cosmic fields are highly non-Gaussian, such that two-point statistics capture only a limited fraction of the available information. While higher-order statistics, such as peak counts, Minkowski functionals, and wavelet moments~\citep{2010MNRAS.402.1049D, 2018MNRAS.474.1116S, 2012PhRvD..85j3513K, 2013PhRvD..88l3002P, 2020MNRAS.499.5902C, 2025JCAP...01..006C}, are more sensitive to nonlinear structures, only a limited subset admit theoretical models that remain sufficiently accurate for cosmological inference in the presence of nonlinear evolution, baryonic uncertainties, and realistic observational systematics. This limitation significantly restricts their applicability in precision cosmology.
\par
These challenges have motivated a shift toward field-level inference (FLI)~\citep{leclercq2025field}, which operates directly on high-dimensional cosmic fields rather than compressed summary statistics. By working with the full fields, FLI naturally preserves non-Gaussian information and avoids reliance on approximate analytic descriptions of nonlinear structure. Instead, it leverages explicit forward modeling to simulate cosmological fields and observational effects, including survey masking, noise, and instrumental systematics, thereby providing a flexible and self-consistent framework for handling complex data models~\citep{2021MNRAS.506L..85L,2024MNRAS.527.1244S}. However, these traditional FLI approaches, while statistically optimal in principle, face several practical limitations. Most notably, they operate in an extremely high-dimensional space, as the latent field typically contains millions of degrees of freedom, making posterior exploration computationally expensive and technically challenging. This high dimensionality places strong demands on sampling algorithms and often leads to scalability issues. In addition, traditional FLI relies on accurate and differentiable forward models that must capture nonlinear gravitational evolution as well as observational systematics; any mismatch or modeling inaccuracy can introduce biases in the inferred parameters.
\par
Machine learning (ML) techniques are increasingly applied in cosmological data processing and analysis~\citep{2015MNRAS.450.1441D, 2017MNRAS.472.1129P, 2017MNRAS.467L.110S, 2024A&A...683A.209Z, 2025NatAs...9..608L, 2025RAA....25l5003W, 2025ApJS..277...12Z}. In cosmological FLI, ML can learn the complex nonlinear mapping between cosmological parameters and high-dimensional fields directly from simulations, greatly reducing the computational cost of traditional FLI~\citep{2019PhRvD.100f3514F, 2020PhRvD.102j3509H, 2024PhRvD.110f3531M}. Furthermore, building on the ML-driven framework, simulation-based inference (SBI) offers a likelihood-free approach that learns the full posterior distribution of cosmological parameters from simulated observables~\citep{2020PNAS..11730055C, 2023MLS&T...4aLT01L, 2025A&A...699A.174S, 2025A&A...699A.327Z}. By leveraging the forward model, SBI naturally incorporates nonlinear evolution and observational systematics, overcoming the limitations of Gaussian-likelihood assumptions and the inefficiency of MCMC sampling in high-dimensional spaces. Consequently, ML-driven SBI has emerged as a scalable and accurate methodology for extracting cosmological information from Stage-IV WL surveys.
\par
In WL cosmology, previous studies have primarily extracted cosmological information from the convergence field~\citep{2013PhRvD..88l3002P, 2018MNRAS.474.1116S, PhysRevD.105.083518, 2023MNRAS.521.2050L} by analyzing its morphology, topology, and peak statistics~\citep{2016PhRvD..94d3533L}. However, the convergence field is not directly observable and must be reconstructed from the measured shear, a process sensitive to the reconstruction algorithm and often degraded by complex survey masks. Additionally, shear measurements involve multiple calibration and correction steps that can introduce systematic uncertainties propagating into the convergence field. To mitigate these effects, this work applies FLI directly to shear fields, while also comparing results with convergence-based inference.
\par
In shear measurements, the dominant noise arises from the intrinsic shapes of galaxies. While intrinsic alignments (IA) induce correlations, these are largely confined to galaxies within the same redshift slice. As a result, the shear signal varies smoothly with redshift, whereas shape noise exhibits stochastic, non-smooth fluctuations. To exploit this property, we perform principal component analysis (PCA) on noisy shear maps across multiple redshift slices: the smooth, coherent components capture the underlying shear signal, while the rapidly fluctuating components correspond to random shape noise. This denoising procedure requires no explicit modeling, analogous to blind-analysis methods commonly used in 21 cm intensity mapping~\citep{2015MNRAS.447..400A}.
\par
In this work, we develop a shear-to-cosmology framework, enabling more direct, information-efficient, and statistically optimal use of the data, particularly in the presence of complex survey geometries. Section~\ref{sec: methodology} presents the methodological framework, including the FLI network and SBI strategy. Section~\ref{sec: mock pipeline} describes the mock pipeline for generating shear fields. Section~\ref{sec: data preprocessing} details shear denoising and convergence reconstruction. Section~\ref{sec: results} presents the inference results, comparing shear- and convergence-based estimates and demonstrating the improvements from PCA denoising. Section~\ref{sec: conclusion} summarizes the main findings.

\section{Methodology}
\label{sec: methodology}

\begin{figure*}
\centering
\includegraphics[width=0.85\textwidth]{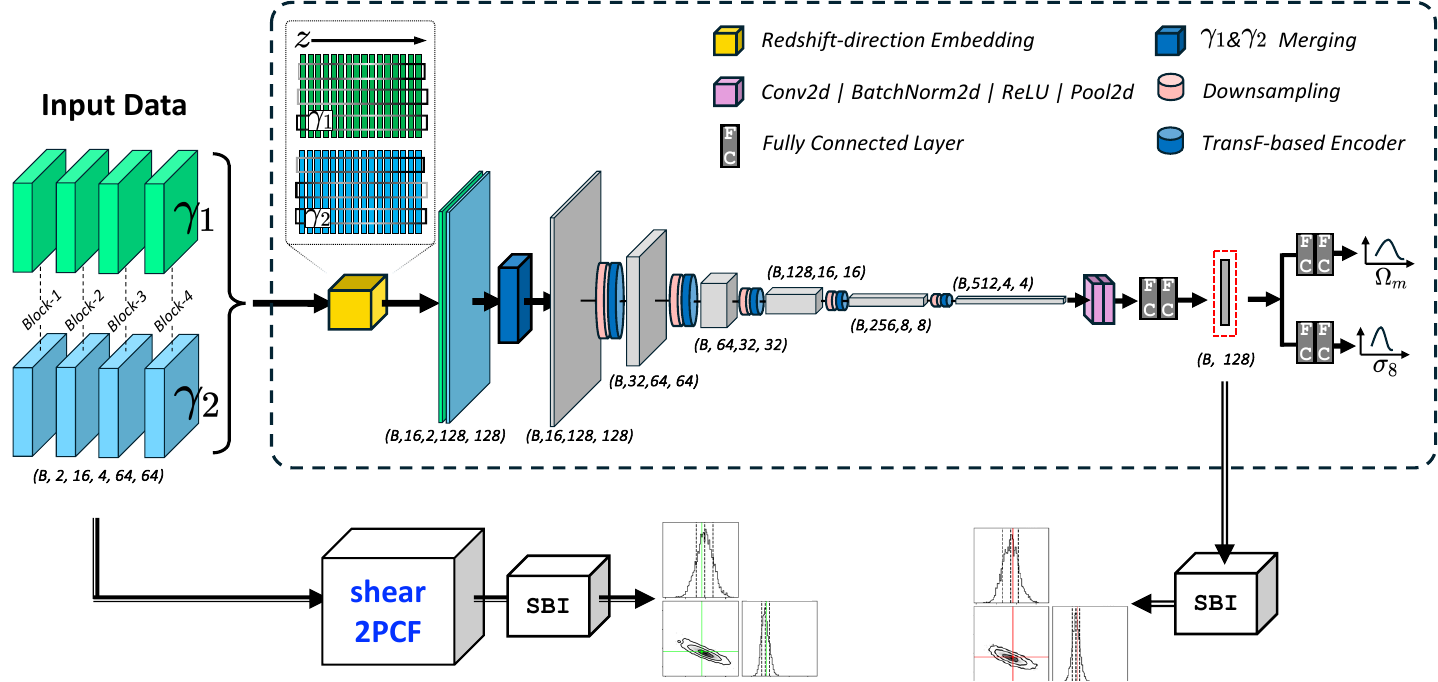}
\caption{\label{fig:FLI-framework} Schematic overview of the proposed framework, consisting of: \textbf{1.} a field-level inference (FLI) network (inside the dashed box) and \textbf{2.} a simulation-based inference (SBI) module (outside the dashed box). The field-level shear input is a six-dimensional tensor with shape $(B,\,N_{\gamma},\,N_{z},\,N_{\rm blk},\,H,\,W)$ $=$ $(B,\,2,\,16,\,4,\,64,\,64)$, where $B$ is the batch size, $N_{\gamma}$ the shear components, $N_{z}$ the number of photo-$z$ slices, $N_{\rm blk}$ the number of blocks in each sky group (see Figure~\ref{fig:sky-108-block-dataset}), and $(H,W)$ the spatial map size. The FLI network comprises four components: (a) a redshift-direction embedding layer, (b) a $\gamma_1$ \& $\gamma_2$ merging layer, (c) Transformer ($\texttt{TransF}$)-based encoders, and (d) fully connected inference layers to predict  $(\Omega_m,\,\sigma_8)$. And the gray boxes show the change of data shapes in the network pipeline.} In parallel, SBI is applied either to the ML-derived feature representation (red dashed box) or to conventional two-point statistics (2PCF). The same structure of FLI network is applied to convergence map $(\kappa)$ by duplicating it to form a two-component input $(\kappa,\kappa)$.
\end{figure*}

The hybrid inference framework, illustrated in Figure~\ref{fig:FLI-framework}, consists of two main components:
\begin{itemize}[leftmargin=4em]
    \item[\textbf{1.}] a FLI network (inside the dashed box), trained to map WL shear fields to cosmological parameters, for generating informative feature representations;
    \item[\textbf{2.}] a SBI module (outside the dashed box) that estimates posteriors from summary statistics, either using conventional shear two-point correlation functions (2PCFs) or ML-derived features.
\end{itemize}

\subsection{Field-Level Inference Network}
\label{subsec: field-level inference}
FLI refers to statistical methods that extract physical fields and cosmological parameters directly from observed data, with the likelihood defined over the entire discretized field and without compressing the map-level information. Neural network can be considered a realization of FLI: it can encode latent representations while still enabling a near-direct mapping from the physical fields to cosmological parameters.

    \subsubsection{Input Data}
\begin{figure}
    \centering
    \includegraphics[width=0.9\columnwidth]{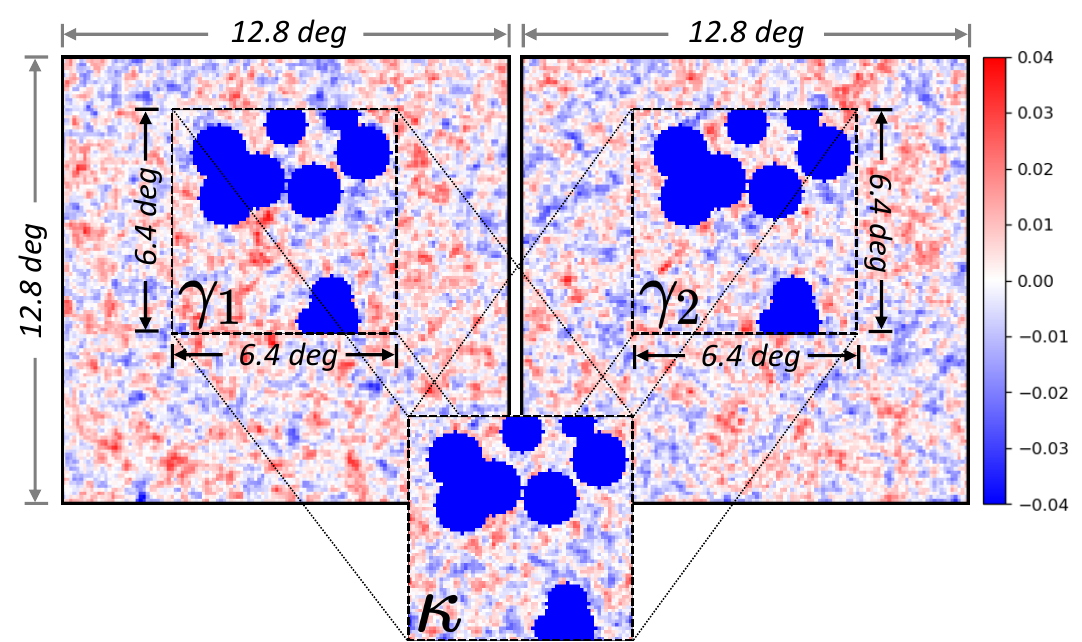}
    \caption{
    \label{fig: input sky block}
    Preparation of an input sky block for training the FLI network. At a resolution of $0.1\,\mathrm{deg/pixel}$, based on the original $12.8\!\times\!12.8\,{\rm deg^2}$ blocks, we generate $6.4\! \times\!6.4\,{\rm deg^2}$ inputs via two augmentations: (1) random cropping (boosting spatial diversity by $64^2$), (2) random masking with overlapping circular masks to mimic survey masks. The corresponding convergence field can then be reconstructed from the prepared shear using the \texttt{KS} algorithm.}
    \end{figure}

    As shown in Figure~\ref{fig:FLI-framework}, the input sample is a six-dimensional shear field, including the 4 sky blocks across 16 photo-$z$ slices. Its shape is $(B,\,N_{\gamma},\,N_{z},\,N_{\rm blk},\,H,\,W)$ $=$ $(B,\,2,\,16,\,4,\,64,\,64)$, where $B$ is the batch size, $N_{\gamma}$ the shear components, $N_{z}$ the photo-$z$ slices, $N_{\rm blk}$ the input sky blocks within each sky group, and $(H,W)$ the spatial map size. Along the redshift direction, the 16 photo-$z$ slices are produced by the denoising procedure (see Section~\ref{subsec: shear denoising}), starting from the 4 photo-$z$ bins of the CSST-like lensing sample (see Section~\ref{subsec: survey setup}) and covering the range $0\!<\!z\!<\!4$. From the sky-region perspective, each input sky block ($6.4\times6.4\,\mathrm{deg}^2$) is constructed using two data-augmentation operations applied to the original $12.8\times12.8\,\mathrm{deg}^2$ sky blocks with a resolution of $0.1\,\mathrm{deg/pixel}$ (see Figure~\ref{fig:sky-108-block-dataset}). The first operation is random cropping, which increases the spatial diversity of the training samples by a factor of $64^2$. The second is random masking, where multiple circular masks with varying radii, allowed to overlap, are applied to mimic realistic survey masks (as illustrated in Figure~\ref{fig: input sky block}). For convergence-based inference, the corresponding convergence maps are subsequently reconstructed from the prepared shear fields using the \texttt{KS} algorithm (section~\ref{subsec: convergence mapping}).
    
    \subsubsection{Network Structure}
    The proposed FLI network (dashed box in Figure~\ref{fig:FLI-framework}) is composed of four key components that collaboratively extract cosmological information directly from the WL shear fields.
    \begin{itemize}[leftmargin=4em]
    \item[\textbf{a.}] \textbf{Redshift-direction Embedding Layer:} 
    The 3D convolutional layers with kernels elongated along the redshift axis capture the shear evolution across photo-$z$ slices.
    \item[\textbf{b.}] \textbf{$\gamma_1\,\&\,\gamma_2$ Merging Layer:} 
    A Transformer block, utilizing the multi-head attention mechanism, is employed to integrate the two shear components.
    \item[\textbf{c.}] \textbf{Transformer-based Encoders:} 
    Unlike conventional convolutional encoders, the Transformer ($\texttt{TransF}$)-based layer captures both local and global dependencies, naturally integrating multi-scale information across redshift slices and sky positions, thereby producing a more informative latent space for cosmological inference.
    \item[\textbf{d.}] \textbf{Inference layers:} 
    For each cosmological parameter, the inference layer consists of two fully connected layers predicting the posterior mean and log-variance. The layers are intentionally kept shallow to facilitate the subsequent SBI in decoding the latent vectors.
    \end{itemize} 
    Together, these layers enable a direct and flexible mapping from observed shear fields to cosmological parameters.
    \par 
    To assess the impact of convergence reconstruction on cosmological inference, we apply the shear-based FLI network to convergence data. Since the network expects two-component inputs, we duplicate the convergence map to form a compatible two-component input ($\kappa$, $\kappa$). This setup enables a direct comparison between shear- and convergence-based inputs while keeping the network architecture and training procedure identical, isolating the effect of input type on inference performance and providing insight into the information loss in reconstructed convergence maps relative to the original shear maps.

    \subsubsection{Loss Function}
    \label{subsubsec: loss function}
    
    Loss function plays a central role in guiding the learning of compression statistics in neural networks, and fundamentally determines the type and structure of the information extracted by the model. 
    Ref.~\cite{2018MNRAS.476L..60A} discusses theoretical approaches to optimal data compression, which, in principle, are capable of preserving the full Fisher information content of the data. However, in practical cosmological data analysis, such optimal compression schemes often require more computational resources, thereby increasing the training time of the network. 
    Ref.~\cite{2025A&A...697A.162L} has shown that different loss function formulations implicitly constrain the types of features learned by the network. 
    
    For instance, a regression-based objective using mean squared error (MSE) encourages the network to learn the conditional mean of the parameters, but does not guarantee that the learned representation is a sufficient statistic for non-Gaussian posteriors. By further introducing a regularization term, such as the Kullback-Leibler (KL) divergence, the latent representation can be constrained to approximate a prescribed prior distribution, thereby improving the structure and generalization of the latent space and mitigating overfitting to some extent.
    \par
    Motivated by this, the proposed FLI network is designed to directly model the posteriors of the cosmological parameters by predicting the means $\mu_i$, log-variances $\log \sigma_i^2$, and parameter samples $\theta_i$ obtained via the reparameterization trick. We then adopt a joint loss function that combines a mean squared error (MSE) term with a KL divergence term, defined as
    \begin{align} 
    L_{\text{tot}} \!=\! \sum_{i=1}^{N_{\rm \theta}} \left( \underbrace{ \left( \theta_i - \hat{\theta}_i \right)^2}_{L_{\rm MSE}} + \underbrace{\frac{\beta}{2} \left(\mu_i^2 + \sigma_i^2 -\log \sigma_i^2 -1\right) }_{L_{\rm KL}} \right) 
    \end{align}
    where $\hat{\theta}_i$ denotes the ground-truth cosmological parameters, and $N_{\theta}$ is the dimensionality of the target parameter space.
    \par 
    During the training process, the validation set is strictly excluded from all parameter updates. It is used solely to monitor the loss and automatically adjust the learning rate when the validation loss plateaus or worsens. Consequently, the evolution of the validation loss provides a direct indication of the model’s performance and training progress. Notably, to evaluate both the shear-based and convergence-based datasets, we employed the same network architecture as mentioned in section~\ref{subsec: field-level inference}.

\subsection{Simulation-Based Inference}
\label{subsec: SBI}

Simulation-based inference (SBI) is a Bayesian framework for parameter estimation when the likelihood is intractable but forward simulations are available. It trains neural density estimators (NDE) on simulated parameter–observation pairs to approximate the posterior. The workflow typically first compresses the forward-modeling maps or summary statistics into a low-dimensional feature space using a neural network, and then trains a conditional density model, often a normalizing flow, to relate these features to the target parameters. Once trained, the model provides an amortized posterior estimator applicable to any (mock) observation for efficient inference. 

We use the \texttt{sbi}\footnote{\url{https://sbi-dev.github.io/sbi}} package~\citep{tejero-cantero2020sbi, boelts2024sbi} and adopt Neural Posterior Estimation (NPE)~\citep{NIPS2016_6aca9700}, specifically the NPE-C (SNPE-C) variant. The goal of NPE is to directly learn an approximation of the posterior distribution $p(\theta\!\mid\!x)$ from a set of simulated parameter–data pairs $(\theta,\;x)$. This is achieved by training a conditional NDE $q_\phi(\theta \!\mid\! x)$ to model the distribution of parameters conditioned on observations. The posterior is modeled with a conditional normalizing flow, implemented as a Neural Spline Flow (NSF), which provides a flexible parameterization of complex posterior distributions. The training relies solely on samples $(\theta,\;x) \sim  p(\theta, x)$, obtained by first drawing parameters from the prior and then generating corresponding simulations through the forward model.

The network parameters $\phi$ are optimized by maximizing the conditional likelihood of the simulated parameters given the data, which is equivalent to minimizing the expected negative log-likelihood over the joint distribution. This objective can be interpreted as minimizing the expected Kullback-Leibler divergence $D_{\mathrm{KL}}$ between the true posterior $p(\theta\,\mid\,x)$ and its neural approximation $q_\phi(\theta \!\mid\! x)$, ensuring that the learned model provides an accurate approximation of the posterior for all simulated pairs. The $D_{\mathrm{KL}}$ is non-negative and vanishes if and only if the two distributions are identical, which leads to the loss function
\begin{equation}
\mathcal{L}(\phi) = -\mathbb{E}_{p(\theta, x)} \left[ \log q_\phi(\theta \!\mid\! x) \right].
\end{equation}

\par
In this paper, from the perspective of SBI, the FLI module (Figure~\ref{fig:FLI-framework}) can be interpreted as a data-compression process that encodes the full WL shear field into a low-dimensional but fully informative summary statistics. Another popular summary statistic for the shear field is the shear-shear correlations. In the following sections, we compare the constraining power of these two types of summary statistics and demonstrate the significant improvements achieved through ML-based compression.

    \subsubsection{Summary Statistics: ML-derived Feature Representations}
    As illustrated in Figure~\ref{fig:FLI-framework}, the neural network compresses each shear map into a 128-dimensional feature representation, denoted as $\bm{x}_{\texttt{ML}}$. The extracted features depend on both the network architecture and the training strategy. Guided by the design of the loss function, the neural compressor is trained to produce informative summary statistics for cosmological parameter inference. Although this objective does not formally guarantee sufficient statistics, it has been shown in practice to retain a substantial fraction of the cosmological information relevant for parameter estimation. For the convergence-based network, we adopt the same architecture as in the shear case but train it independently, enabling the network to transform each convergence map into a representation of the same length (128).
    
    \subsubsection{Summary Statistics: Shear 2PCFs}
    In the traditional statistics of cross-$z$ 2PCFs, the shear data originally consist of 16 photo-$z$ slices. To reduce the computational cost, adjacent slices are averaged to produce 8 $z$-slices, resulting in 36 distinct pairs. For each pair $(a,b)$, the correlation functions are defined as
    \begin{align}
        \xi_{+}^{(a,b)}(\vartheta) &= \left\langle \gamma^{a}_{\rm t} \,\gamma^{b}_{\rm t} \right\rangle (\vartheta) + \left\langle \gamma^{a}_{\times} \, \gamma^{b}_{\times} \right\rangle (\vartheta), \\
        \xi_{-}^{(a,b)}(\vartheta) &= \left\langle \gamma^{a}_{\rm t} \, \gamma^{b}_{\rm t} \right\rangle (\vartheta) - \left\langle \gamma^{a}_{\times} \, \gamma^{b}_{\times} \right\rangle (\vartheta), 
    \end{align}
    where $\gamma_{\rm t}$ and $\gamma_{\times}$ denote the tangential and cross components of the shear, respectively. Based on  $\texttt{TreeCorr}$\footnote{\url{https://rmjarvis.github.io/TreeCorr}}~\citep{2004MNRAS.352..338J,2015ascl.soft08007J} library, the measurements are performed over angular scales in the range $\vartheta_{\rm min}=6$ to $\vartheta_{\rm max}=120$, sampled at $N_{\vartheta}=5$ logarithmically spaced points. To combine the information from all $z$-slice pairs, we first stack the $\xi_{+}$ and $\xi_{-}$ measurements across different pairs, 
    \begin{align}
        \hat{\xi}_{+} &= \texttt{stack(}\, \xi_{+}^{p1},\xi_{+}^{p2},\,\cdots, \,\xi_{+}^{p36} \,\texttt{)}, \\ 
        \hat{\xi}_{-} &= \texttt{stack(}\, \xi_{-}^{p1},\xi_{-}^{p2},\,\cdots, \,\xi_{-}^{p36} \,\texttt{)},
    \end{align}
    and subsequently concatenate the two to construct the summary statistics, $\hat{\xi}_{+-} = \texttt{stack(} \hat{\xi}_{+}, \hat{\xi}_{-}\texttt{)}$ with length of 360.

    Although the traditional statistics ($\hat{\xi}_{+-}$) involve larger data volumes than ML-derived features, subsequent SBI results show that the latter capture richer cosmological information and yield tighter parameter constraints.

\section{Mock Pipeline}
\label{sec: mock pipeline}

\subsection{N-body Simulation}
\label{subsec: n-body simulation}

To generate weak lensing maps for training and validation, we use the publicly available $\mathrm{C\text{\scriptsize OSMO}G\text{\scriptsize RID}V1}$\footnote{\url{http://www.cosmogrid.ai}} simulation suite~\citep{2209.04662,PhysRevD.105.083518}, 
which provides a diverse set of cosmologies suitable for Stage-IV lensing survey conditions. The mock catalogs are constructed to match the typical redshift distribution and galaxy number density expected in forthcoming wide-field lensing surveys (e.g., CSST, Euclid, Roman).
\par
$\mathrm{C\text{\scriptsize OSMO}G\text{\scriptsize RID}V1}$ contains 2,500 cosmologies sampled within the $w$CDM framework, varying ${\Omega_m,\,\sigma_8\,,w_0,\,n_s,\,\Omega_b,\,h_0}$, and assuming three degenerate neutrinos with a fixed mass sum $\sum m_{\nu}=0.06\,{\rm eV}$. For each cosmology, seven realizations with different initial conditions were produced. The full suite includes cosmologies drawn from both wide- and narrow-grid priors. In this work, we use only the “$\texttt{run0}$” realization of 1,260 cosmologies sampled from the narrow-grid prior, spanning the parameter ranges $\Omega_m\in[0.15,0.45]$, $\sigma_8\in[0.5,1.3]$, $w_0\in[-1.25,-0.75]$, $n_s\in[0.93,1.00]$, $\Omega_b\in[0.04,0.05]$, and $h_0\in[0.65,0.75]$. Based on estimates from two-point statistics ($\partial C_\ell / \partial \theta(\ell)$, \citep{2024MNRAS.527.5206Y}, the four parameters of $(n_s, h_0, w_0, \Omega_b)$ highly depend on angular sensitivity across over two orders of angular scales, which is limited by our ``$\rm 6\,arcmin$" small-scale cut due to the simulation limit. Therefore, as a proof of methodology, we focus on $(\Omega_m,\,\sigma_8)$ to represent the cosmological constraining power in this work.
\par
Based on the DM simulation snapshots, $\mathrm{C\text{\scriptsize OSMO}G\text{\scriptsize RID}V1}$ also provides projected full-sky DM lightcone shells covering the redshift range $0\!<\!z\!<\!4$, stored in HEALPix format with resolutions of $N_{\mathrm{side}}=512$ or 1024. Owing to computational limitations, we use the $N_{\mathrm{side}}\!=\!512$ shells as the basis for the ray-tracing procedure to generate the weak lensing signals in this work.

\subsection{Ray-tracing}
\label{subsec: Ray-tracing}
Utilizing the $\sim\!68$ redshift shells provided in the $\mathrm{C\text{\scriptsize OSMO}G\text{\scriptsize RID}V1}$ data covering the redshift interval of $z\in [0,4]$,  we employ the full-sky ray-tracing algorithm $\texttt{DORIAN}$\footnote{\url{https://gitlab.mpcdf.mpg.de/fferlito/dorian}} ~\citep{2024MNRAS.533.3209F} to generate WL shear signals at the followed 23 redshifts: $z$ $\in$ $\{0.20,$ $0.30,$ $0.40,$ $0.50,$ $0.60,$ $0.70,$ $0.80,$ $0.90,$ $1.00,$ $1.10,$ $1.20,$ $1.30,$ $1.40,$ $1.50,$ $1.60,$ $1.70,$ $1.80,$ $2.00,$ $2.20,$ $2.40,$ $2.80,$ $3.20,$ $3.60\}$.
\par
Notably, the simulation timesteps are not fully synchronized across different cosmologies, leading to discrepancies in the redshift sampling of the DM lightcone shells, especially at high redshifts. Such inconsistencies can introduce systematic errors and spurious features into the machine learning dataset. To address this, we adopt a coarser redshift sampling for the shear fields in the high-redshift regime, which both mitigates potential systematics and better matches the expected galaxy distributions in WL surveys, while substantially reducing computational costs.

\begin{center}
\includegraphics[width=0.85\columnwidth]{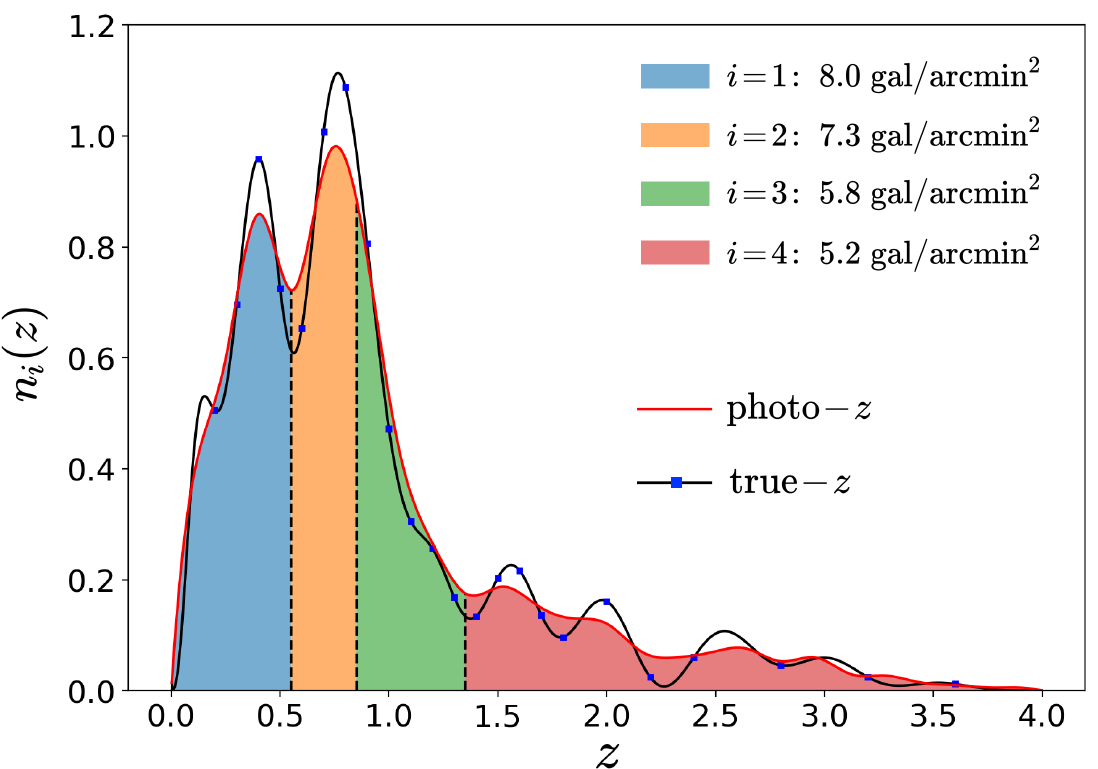}
\captionof{figure}{
CSST-like photo-$z$ distribution. The red curve shows the photo-$z$ distribution with photo-$z$ uncertainties $\sigma(z)=\sigma_z(1+z)$, $\sigma_z=0.05$. Using this $z$-dependent kernel, the true-$z$ distribution (marked curve) is obtained by deconvolution. A mock true-$z$ galaxy catalog is generated assuming a galaxy number density of $26\,{\rm gal/arcmin^2}$ and intrinsic shape noise $\sigma_e = 0.288$. Incorporating photo-$z$ uncertainties, the mock photo-$z$ catalog is then constructed. To enhance the S/N of the WL signal, the photo-$z$ catalog is divided into 4 $z$-bins, each shown in a different color with its corresponding galaxy number density.}
\label{fig:CSST photo-z distribution} 
\end{center}

\subsection{Survey Setup}
\label{subsec: survey setup}

To validate the hybrid framework under realistic conditions, we employ mock photo-$z$ galaxy catalogs with specifications comparable to high-precision surveys such as CSST. These mocks include galaxy shape noise and irregular survey masking to emulate the observational conditions expected in large-scale weak-lensing programs.
\par
WL shear measurements rely on accurately determining galaxy ellipticities. In theory, the observed ellipticity ($\varepsilon$) of a galaxy can be approximated as a linear combination of its intrinsic ellipticity ($\varepsilon^{s}$) and the lensing shear ($\gamma$), given by 
\begin{align}
\label{eq: add shape noise}
\varepsilon = \frac{\varepsilon^{s} + \gamma}{1 + \varepsilon^{s} \gamma^{*}}
\end{align}
where $\gamma = \gamma_1 + \gamma_2\,{\rm i}$,  $\varepsilon^{s}=\varepsilon^{s}_1 + \varepsilon^{s}_2\,{\rm i}$ and ``${\,\rm i \,}$" denotes the imaginary unit. 
It is assumed that the two components of $\varepsilon^{s}$ follow a Gaussian distribution with standard deviation $\sigma_e = 0.288$, consistent with the value adopted in the KiDS survey \citep{2020A&A...633A..69H, 2020A&A...642A.158B}. Ideally, the ensemble average of galaxy ellipticities can provide an unbiased shear estimator, with $\langle \varepsilon^{s} \rangle = 0$ and $\gamma = \langle \varepsilon \rangle$. In practice, however, the accuracy is limited by the finite surface density of usable source galaxies: insufficient samples leave residual contributions from intrinsic shapes, introducing galaxy shape noise into the WL signal. The third-generation surveys typically achieve galaxy number densities of approximately 10–20 $\rm gal/arcmin^2$, whereas upcoming or ongoing fourth-generation surveys are expected to reach densities as high as about 30 $\rm gal/arcmin^2$. China Space Station Survey Telescope (CSST) is a Stage-IV space-based observatory for wide-field photometric imaging and slitless spectroscopy, targeting a sky coverage of $17{,}500\,\mathrm{deg}^2$~\citep{2018MNRAS.480.2178C, 2019ApJ...883..203G, zhan2021wide}. Its upcoming WL survey is expected to achieve a galaxy number density of $\sim\! 26\,\mathrm{gal/arcmin^2}$~\citep{2019ApJ...883..203G, 2024SCPMA..6770413L}.

Cropping shear maps from full-sky HEALPix maps at different redshifts yields a series of pure multi-redshift shear maps. To generate the mock WL catalog, galaxies are placed at the centers of the corresponding mesh pixels, ignoring their finer spatial distribution within each pixel. Based on the photometric-redshift (photo-$z$) distribution of CSST lensing samples~\citep{2024MNRAS.527.5206Y}, a CSST-like catalog can then be generated through the following steps.
\begin{itemize}[leftmargin=4em]
    \item[(1)] We adopt the photo-$z$ distribution of CSST lensing galaxies (red curve in Figure~\ref{fig:CSST photo-z distribution}), assuming a photo-$z$ uncertainty parameterized as $\sigma(z) = \sigma_z (1+z)$, with $\sigma_z=0.05$~\citep{2018MNRAS.480.2178C}. The corresponding true-$z$ distribution is then recovered through a deconvolution procedure.
    \item[(2)] Using the recovered true-$z$ distribution (black curve in Figure~\ref{fig:CSST photo-z distribution}) and an assumed galaxy number density of $26~\mathrm{gal/arcmin^2}$, we generate a mock sample of galaxies with redshifts drawn accordingly.
    \item[(3)] Finally, by assigning shear signals from simulated shear fields at multiple redshifts (blue squares in Figure~\ref{fig:CSST photo-z distribution}) and incorporating intrinsic shape noise with a dispersion of $\sigma_e = 0.288$, we construct the CSST-like photo-$z$ galaxy catalog. 
\end{itemize}

As a preliminary investigation within the shear-to-cosmology framework, this work focuses solely on extracting cosmological information in the presence of galaxy shape noise and photo-$z$ uncertainties, deferring the inclusion of PSF effects, shear measurement biases, mean redshift biases, intrinsic alignments (IA), baryonic feedback, reduced shear, source clustering, magnification, and other systematics to future studies using larger simulation sets.

\section{Data Preprocessing}
\label{sec: data preprocessing}

\begin{figure*}
\centering
\includegraphics[width=0.85\textwidth]{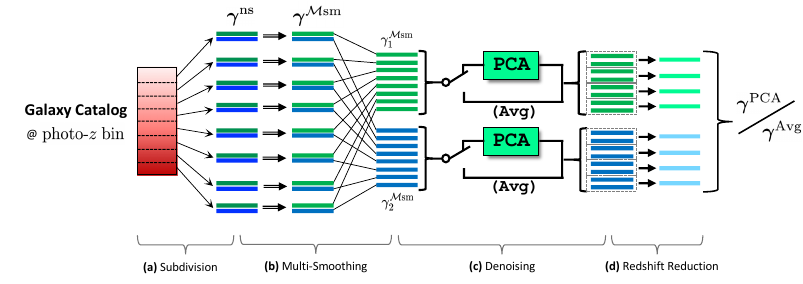}
\caption{\label{fig: shear-denoisng} 
Pipeline of shear measurement in each photo-$z$ bin including 4 steps to enhance the shear signal quality. When PCA processing is applied, the denoised shear field is denoted as $\bm{\gamma}^{\rm PCA}$; otherwise, the resulting field is denoted as $\bm{\gamma}^{\rm Avg}$.}
\end{figure*}

The goal of employing FLI is to maximally extract the non-Gaussian information encoded in the data. However, conventional smoothing operations inevitably suppress such features, thereby motivating the development of denoising techniques tailored for FLI. These methods aim to reduce noise while preserving the essential non-Gaussian structures of the field.
\par 
From an information-theoretic perspective, denoising does not introduce additional cosmological information into the data. Instead, it facilitates more efficient information extraction by improving the pixel-level signal-to-noise (S/N) ratio. In practice, ML models operate under a trade-off between effective information extraction and the control of overfitting, where limited data and finite model capacity can hinder the full utilization of the available information. By enhancing the S/N ratio, denoised data enable models to more reliably capture the underlying cosmological signal, thereby improving the fidelity and constraining power of field-level inference.
\par 
Several ML-based denoising approaches, such as GAN-based models~\citep{2021MNRAS.504.1825S} and diffusion models~\citep{2025arXiv250500345A}, have been applied to WL mass maps. In contrast, leveraging the distinct statistical properties of WL signals and shape noise, we aim to develop a blind, training-free denoising technique that does not rely on any learned model. This approach constitutes a key methodological component of this work.

\subsection{Shear Denoising}
\label{subsec: shear denoising}

WL signals are strongly redshift-dependent, whereas shape noise is largely uncorrelated with redshift. This motivates using PCA algorithm along the redshift direction to separate signal from noise. However, the global nature of PCA causes high S/N modes from high-redshift bins to leak into lower-redshift bins when wide redshift bins are used, diluting the true redshift evolution. To mitigate this, we perform PCA independently within narrower redshift intervals, limiting cross-bin signal mixing. This requires subdividing galaxies within each photo-$z$ bin, enabling the denoising procedure to better preserve the redshift dependent structure of the shear signal. Hence, the entire mock galaxy catalog over the entire photo-$z$ range is divided into 4 photo-$z$ bins, as shown in Figure~\ref{fig:CSST photo-z distribution}. Considering both the photo-$z$ distribution of CSST galaxies and the characteristic decline of WL strength at lower redshifts, the number of galaxies in the low photo-$z$ bins is appropriately increased to boost the WL signal in those bins.

\begin{center}
\includegraphics[width=\columnwidth]{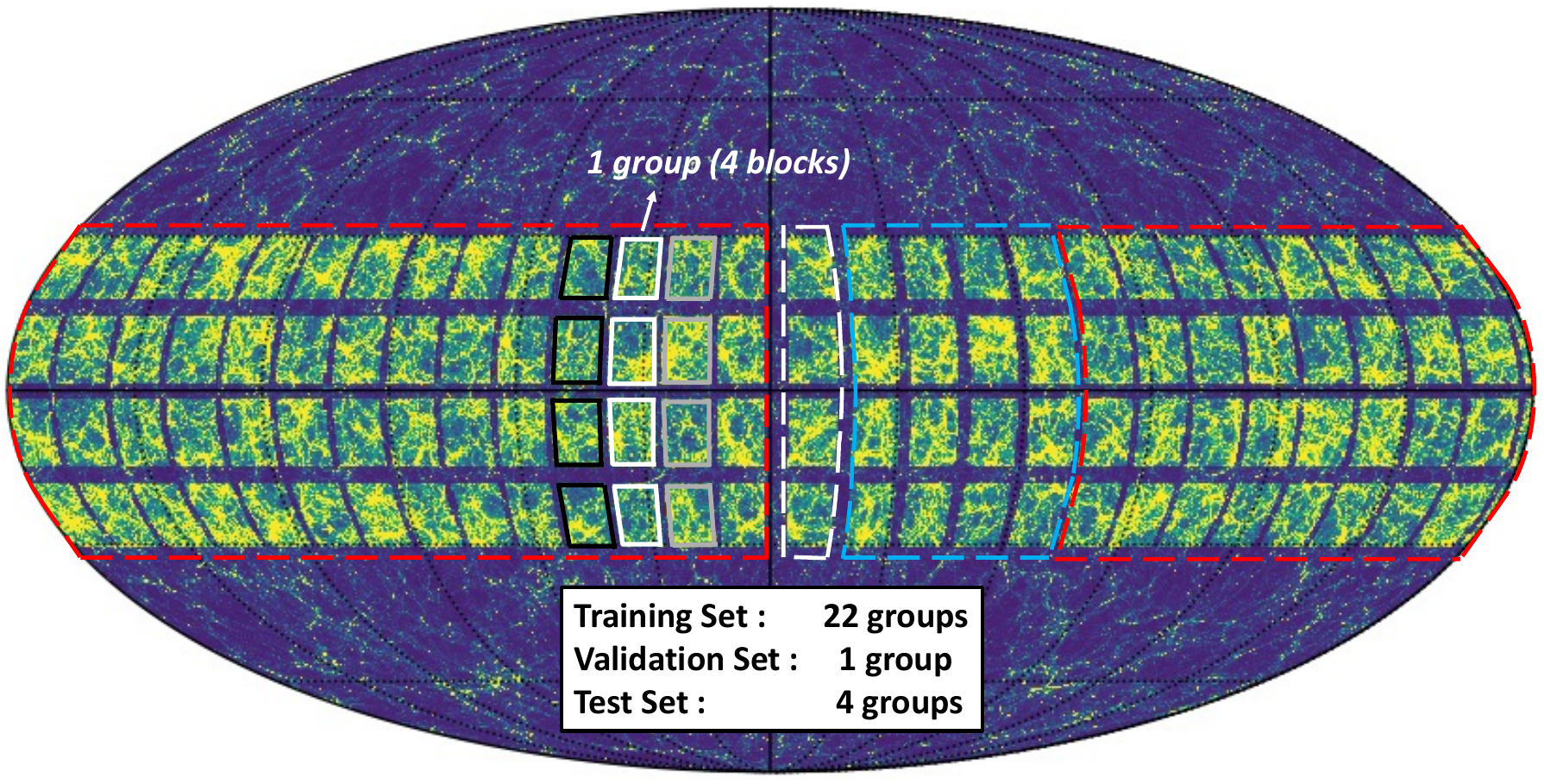}
\captionof{figure}{Illustration of the sky-region partitioning into training, validation, and test sets. For each cosmology, the dataset contains 108 sky blocks, grouped in fours within the same declination region. Groups enclosed by red, blue, and white dashed boxes indicate the training, test, and validation sets, respectively, with no overlapping sky regions. Each block is a $128 \times 128$ mesh spanning $12.8 \times 12.8\,{\rm deg^2}$ and is randomly cropped to $6.4 \times 6.4\,{\rm deg^2}$ before input to the neural network. Each group of cropped blocks forms a sample, covering $163.84\,\mathrm{deg^2}$.}
\label{fig:sky-108-block-dataset}
\end{center}

For each sky block in Figure~\ref{fig:sky-108-block-dataset}, the shear data can be obtained through the following steps (illustrated in Figure~\ref{fig: shear-denoisng}), incorporating analytical operations to suppress noise and enhance the shear signal quality. 
\begin{center}
\includegraphics[width=0.90\columnwidth]{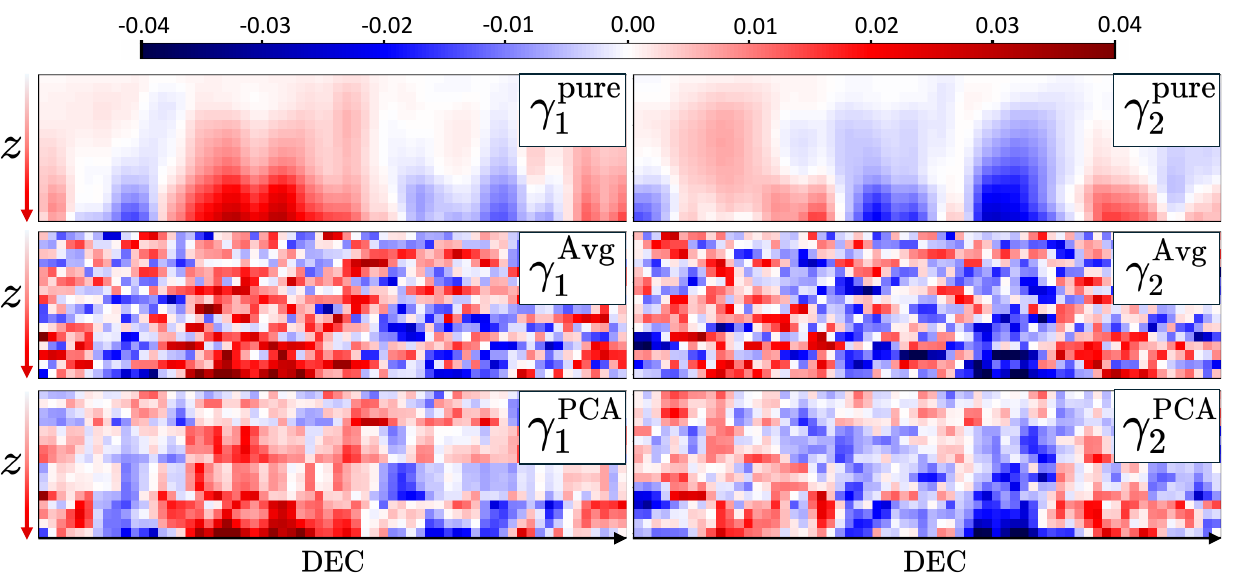}
\captionof{figure}{Visualization of shear denoising. The bottom panels show shear fields from the pipeline in Figure~\ref{fig: shear-denoisng} including PCA denoising ($\bm\gamma^{\rm PCA}$), while the middle panels show shear fields from the pipeline without PCA ($\bm\gamma^{\rm Avg}$). The top panels display the corresponding pure shear signals after the same multi-smoothing operation ($\bm\gamma^{\rm pure}$). In each panel, the horizontal axis corresponds to declination and the vertical axis to redshift. As shown in the figure, PCA denoising clearly preserves more reliable redshift structures, which provide the key information for our inference framework.}
\label{fig: effect of denoising}
\end{center}

\begin{itemize}[leftmargin=4em]
    \item[\textbf{(a)}] $\textbf{Subdivision: }$
    To ensure statistical stability, each subset must contain a sufficient number of galaxies (typically 20-40 per pixel), given the finite sample size within each photo-$z$ bin. At the same time, the number of subsets determines the number of effective PCA modes; using too few subsets reduces the decomposition capacity and limits the ability to separate signal from noise. We therefore adopt 8 subsets as a pragmatic compromise between these competing requirements. Accordingly, within each photo‑$z$ bin and sky block, we divide galaxies into 8 subsets by their photo‑$z$ values, then compute noisy shear maps (${\bm \gamma}^{\rm ns}$) by averaging the ellipticities per subset, expressed as 
        \begin{align}
        {\bm \gamma}^{\rm ns} = \left\langle \bm{\varepsilon} \right\rangle.
        \end{align}
    Notably, we adopt the simplified precondition here that all galaxies reside entirely within pixel boundaries, neglecting the complexities associated with those intersecting pixel edges.
    \item[\textbf{(b)}] $\textbf{Multi-Smoothing: }$
    To improve PCA decomposition by enhancing the S/N ratio via high-frequency noise suppression, we smooth the noisy shear map in each subset using a set of Gaussian kernels,
        \begin{align}
        {\bm \gamma}^{\rm sm} (\bm \theta) &= \left[ {\bm \gamma}^{\rm ns} \otimes \mathcal{G} \right] (\bm \theta) \\ \nonumber  
        &\rightleftharpoons  \;\; \tilde{{\bm \gamma}}^{\rm sm} (\bm \ell) = \tilde{\mathcal{G}}(\bm \ell) \cdot \tilde{{\bm \gamma}}^{\rm ns} (\bm \ell)
        \end{align} 
    where the Gaussian kernel in Fourier space is given by
        \begin{align}
        \tilde{\mathcal{G}}(\bm \ell) \coloneqq {\rm exp}(-\frac{|\bm\ell|^2}{2\tilde{\sigma}_{\rm sm}^2}).
        \end{align} 
    Here we adopt a set of smoothing scales in Fourier space, $\tilde{\sigma}_{\rm sm} = [10,\,20,\,40]$ $\rm pixel^{-1}$, which correspond to Gaussian kernel widths in real space of $\sigma_{\rm sm} = [1.00,\,0.50,\,0.25]$ deg, respectively. 
    Intuitively, feeding multiple independently smoothed results into a neural network preserves multi-scale, non-Gaussian features. Although increased smoothing improves the S/N ratio, it reduces local complexity, making overfitting more likely. Averaging multiple smoothed results may offer a balance between information retention and overfitting prevention, expressed as
        \begin{align}
        {\bm \gamma}^{\msm}(\bm \theta) = \frac{1}{N_{\rm sm}} \sum_{n=1}^{N_{\rm sm}} \,{\bm \gamma}^{\rm sm}(\bm \theta).
        \end{align}
    \item[\textbf{(c)}] $\textbf{Denoising: }$
    Conventional denoising algorithms (like smoothing and Wiener-filter) normally down-weight the high-$\ell$ modes at map-level, while ignoring the fact that shear signals are correlated along the redshift-direction and shape noises are not. Based on the resulting ${\bm \gamma}^{\msm}$ from 8 subsets, one can further suppress the noise level via the denoising algorithm --  Principal Component Analysis ($\texttt{PCA}$): 
        \begin{align}
            \begin{bmatrix}
            {\bm \gamma}^{\rm PCA}_{(1)}\\
            {\bm \gamma}^{\rm PCA}_{(2)}\\
            \cdots \\
            {\bm \gamma}^{\rm PCA}_{(8)}\\
            \end{bmatrix}  
            = \texttt{PCA}\,
            ( 
                \begin{bmatrix}
                {\bm \gamma}^{\msm}_{(1)}\\
                {\bm \gamma}^{\msm}_{(2)}\\
                \cdots \\
                {\bm \gamma}^{\msm}_{(8)}\\
                \end{bmatrix}
            ).
        \end{align}
    Based on the $\texttt{PCA}$ decomposition, the input shear data are separated into 8 orthogonal modes. We retain only the first 2 smooth principal components for shear reconstruction, since the true shear field varies smoothly with redshift, whereas the shape noise manifests as random fluctuations.
    \item[\textbf{(d)}] $\textbf{Redshift Reduction: }$
    Combining the outputs from the 4 photo-$z$ bins yields a four-dimensional shear dataset ($N_{\bm{\gamma}}\!=\!2$, $N_{z}\!=\!32$, $H$, $W$), where $N_{\bm{\gamma}}$ and $N_{z}$ denote the two shear components and the redshift dimension after denoising, respectively. To reduce its data size for ML or statistical analyses, every two consecutive redshift layers are averaged, resulting in ($N_{\bm{\gamma}}\!=\!2$, $N_{z}\!=\!16$, $H$, $W$). The processed data is denoted as $\bm{\gamma}^{\rm PCA}$ when PCA is applied, and $\bm{\gamma}^{\rm Avg}$ otherwise.
\end{itemize}

At present, a systematic sensitivity analysis of these design choices (e.g., the number of subsets and the smoothing scale) is still lacking. Nevertheless, empirical results suggest that the method remains robust to moderate variations, provided the parameters stay within physically and statistically motivated ranges. In practice, the number of subsets is governed by a trade-off between ensuring sufficient galaxy counts per subset and retaining enough degrees of freedom for effective PCA decomposition, while the smoothing scale reflects a balance between noise suppression and the preservation of small-scale structures. As long as these considerations are satisfied, we do not expect qualitative changes in the overall performance.

\par
It is worth noting that we employed a small trick: the mask effect is applied to the prepared shear maps (${\bm \gamma}^{\rm PCA}$ or ${\bm \gamma}^{\rm Avg}$). Although this treatment does not fully correspond to the actual mask, it can significantly accelerate the training of the neural network.

\subsection{Convergence Mapping}
\label{subsec: convergence mapping}

Based on the prepared shear field (${\bm \gamma}^{\rm PCA}$ or ${\bm \gamma}^{\rm Avg}$), generated with random masks during training, we apply the \texttt{KS} algorithm~\citep{1993ApJ...404..441K} to reconstruct the convergence field ($\kappa$), retaining only its E-mode component:
\begin{align}
   \kappa^{\rm prep}(\bm \theta) \gets \texttt{KS}\texttt{(} \,\bm{\gamma}^{\rm prep}(\bm\theta)\,\texttt{)}{\bm :}\quad 
   & \tilde{\kappa}^{\rm prep}(\bm \ell) = D^{*}(\bm \ell)\,\tilde{\bm{\gamma}}^{\rm prep}(\bm\ell),\\ 
   & D(\bm \ell) = \frac{\ell_{1}^2 - \ell_{2}^2 +2\,{\rm i}\,\ell_{1}\ell_{2}}{|\bm \ell|^2}, \nonumber
\end{align}
where $D(\bm\ell)$ is the projection operator, ``$\,\rm i\,$" denotes the imaginary unit, and the option ``$\rm prep$" can be specified as ``${\rm PCA}$" or ``${\rm Avg}$".

\section{RESULTS}
\label{sec: results}
    
\subsection{Loss Evolution}

As illustrated in Figure~\ref{fig: effect of denoising}, PCA denoising effectively enhances the S/N ratio of the shear data and further reduces the training loss. To evaluate its impact, we compare the FLI results with and without PCA denoising. Based on the loss function described in Sec.~\ref{subsubsec: loss function}, Figure~\ref{fig:loss-evolution} presents the training loss evolution with a batch size of 32 for four types of input data. The first two are convergence fields, the averaged convergence ($\kappa^{\rm Avg}$) and the PCA-denoised convergence ($\kappa^{\rm PCA}$), both reconstructed from their corresponding shear fields using the \texttt{KS} algorithm. The other two are shear fields: the averaged shear ($\gamma^{\rm Avg}$) and the PCA-denoised shear ($\gamma^{\rm PCA}$). As shown in Figure~\ref{fig:loss-evolution}, the shear-based approaches achieve substantially lower losses than the convergence-based ones, since convergence reconstruction introduces additional systematic errors, particularly in the presence of sky masking.

\begin{center}
\includegraphics[width=0.95\columnwidth]{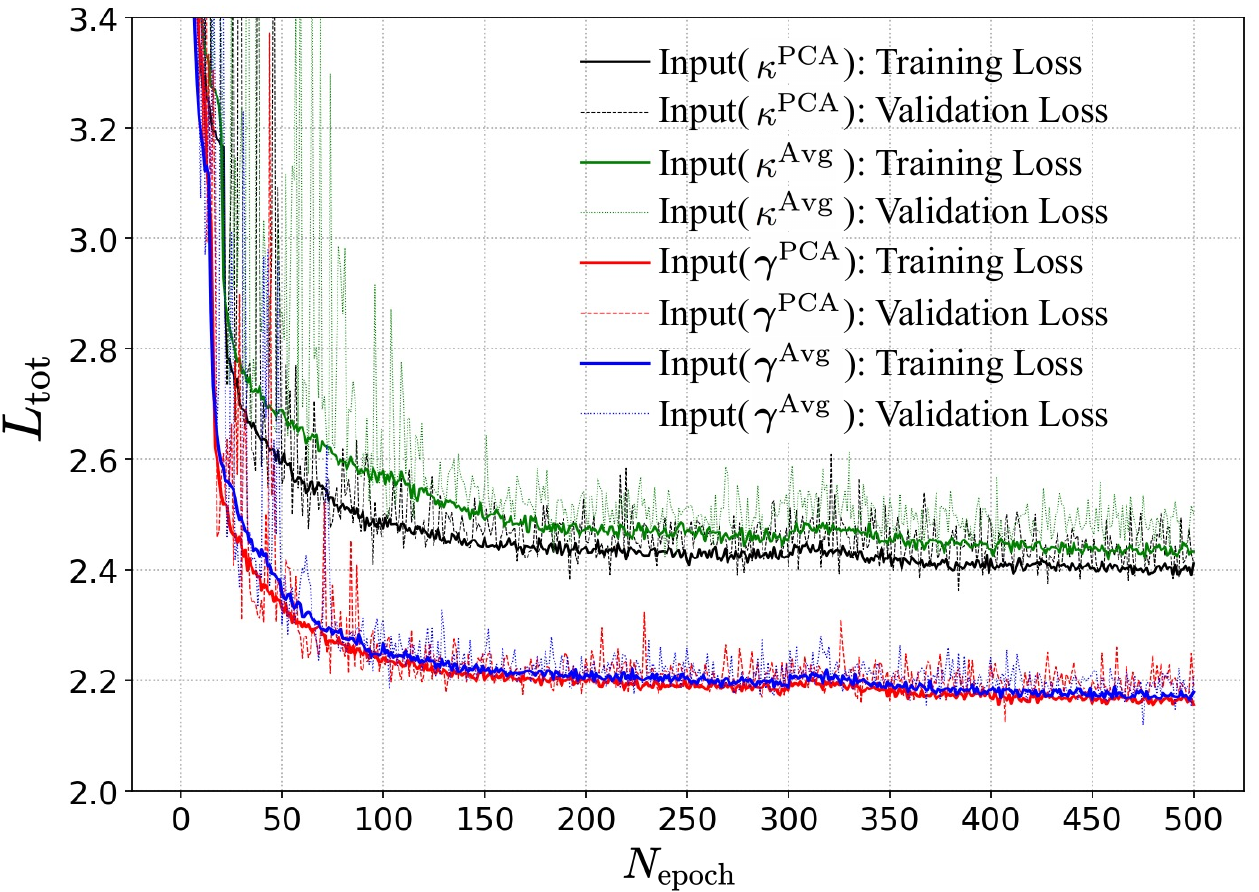}
\captionof{figure}{Evolution of the loss ($\texttt{batch\;size=32}$) for different input data: the shear maps $\bm{\gamma}^{\rm PCA}$ (with denoising) and $\bm{\gamma}^{\rm Avg}$ (without denoising), the corresponding convergence maps $\kappa^{\rm PCA}$ and $\kappa^{\rm Avg}$ via \texttt{KS} algorithm. In neural network training, the validation set is excluded from parameter optimization and serves solely to trigger automatic learning rate reduction based on variations in its loss. Consequently, the evolution of validation loss reflects the model’s performance improvement. }
\label{fig:loss-evolution} 
\end{center}

\begin{figure*}
\centering
\includegraphics[width=0.70\textwidth]{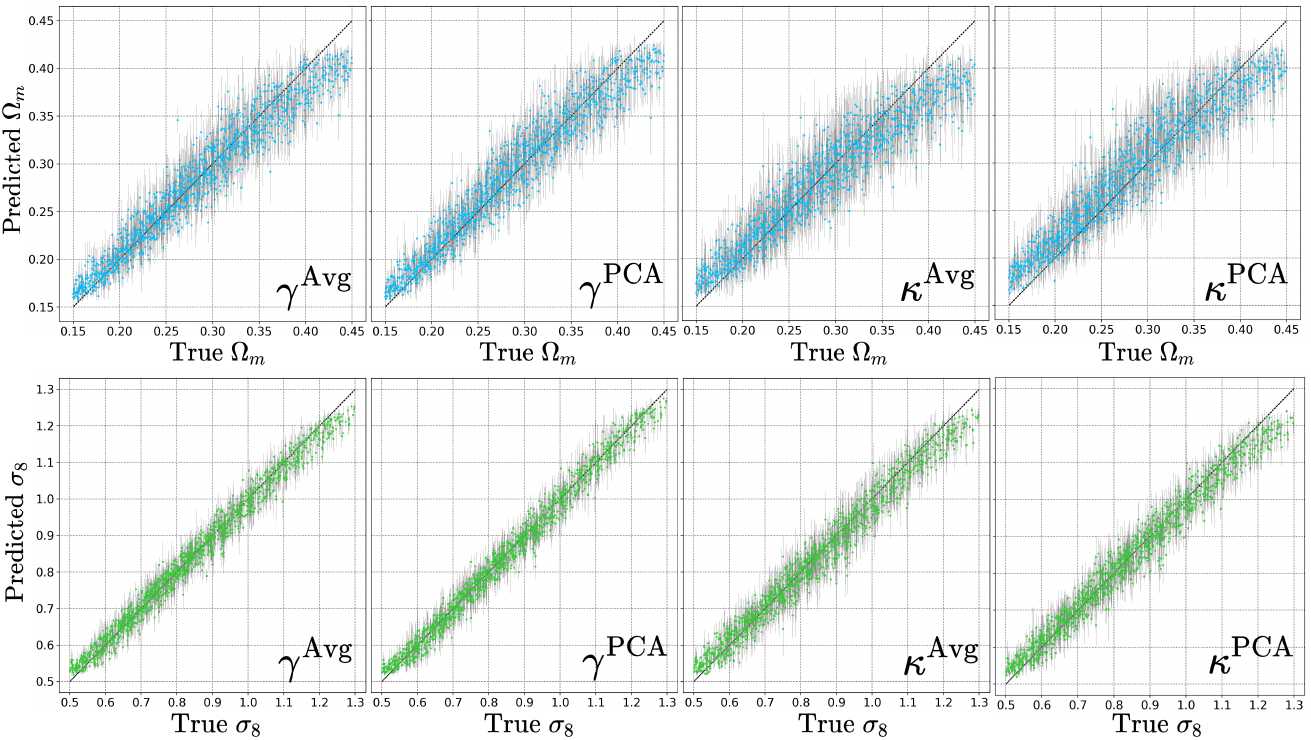}
\caption{\label{fig: Om-s8 predictions}
Predictions of $\Omega_{m}$ and $\sigma_8$ directly from networks trained and tested on four types of input data: (${\bm\gamma}^{\rm Avg}$, ${\bm\gamma}^{\rm PCA}$, $\kappa^{\rm Avg}$, $\kappa^{\rm PCA}$). For both shear- and convergence-based inputs, the networks share the same architecture (Figure~\ref{fig:FLI-framework}) but are trained independently. Each point, evaluated on the test set, denotes the average prediction over 4 sky groups (covering $655.36\,\mathrm{deg}^2$) within the same cosmology, with error bars indicating the standard deviation. Dashed lines mark the true parameter values.}
\end{figure*}

\subsection{Predictions by FLI}
Under the assumption of Gaussian distribution, the trained FLI network can output the posteriors of the target parameters ($\Omega_{m}$, $\sigma_8$) directly. As shown in Figure~\ref{fig:sky-108-block-dataset} in Sec.~\ref{sec: data preprocessing}, the prepared test set includes 4 discrete sky groups per cosmology, covering a total of $655.36\,\mathrm{deg}^2$. In order to highlight the differences between the input data types: (${\bm\gamma}^{\rm Avg}$, ${\bm\gamma}^{\rm PCA}$, $\kappa^{\rm Avg}$, $\kappa^{\rm PCA}$), we average the FLI predictions from the 4 sky groups for each cosmology in Figure~\ref{fig: Om-s8 predictions}. As shown in the figure, qualitatively, shear-based inference outperforms convergence-based inference. Furthermore, applying PCA denoising reduces the systematic bias of the predictions relative to the true values.

\subsection{Validation of Predictions}

Probability Integral Transform (PIT) is a standard diagnostic tool for evaluating the reliability of model-predicted probability distributions. When a model’s cumulative distribution function (CDF) is well calibrated, the PIT values obtained by evaluating observed data against the CDF follow a uniform distribution. PIT thus offers a simple yet effective way to assess both the calibration and uncertainty of probabilistic predictions.
\par 
In this work, cosmological parameters are inferred using two approaches:
(1) directly from the FLI network, and (2) SBI applied to summary statistics. We perform PIT analyses on the posteriors obtained from both methods to evaluate their inference quality and provide an additional consistency check for our framework.

\par
As shown in Figure~\ref{fig: PIT validation}, the PIT distributions from SBI are noticeably closer to uniform than those from the FLI network, indicating better-calibrated posteriors. This is expected because, although both methods use the same ML-derived feature representations, the inference module in the FLI network, comprising fully connected layers (Figure~\ref{fig:FLI-framework}), assumes Gaussian posteriors, which limits flexibility and accuracy. In contrast, SBI directly learns the mapping from features to parameter posteriors using simulated data, enabling it to capture complex, non-Gaussian, and even multi-modal distributions. By explicitly modeling the full posterior, SBI naturally accounts for parameter correlations and uncertainties, yielding more reliable and accurate inference than the Gaussian-constrained network outputs.

\begin{center}
\includegraphics[width=0.7\columnwidth]{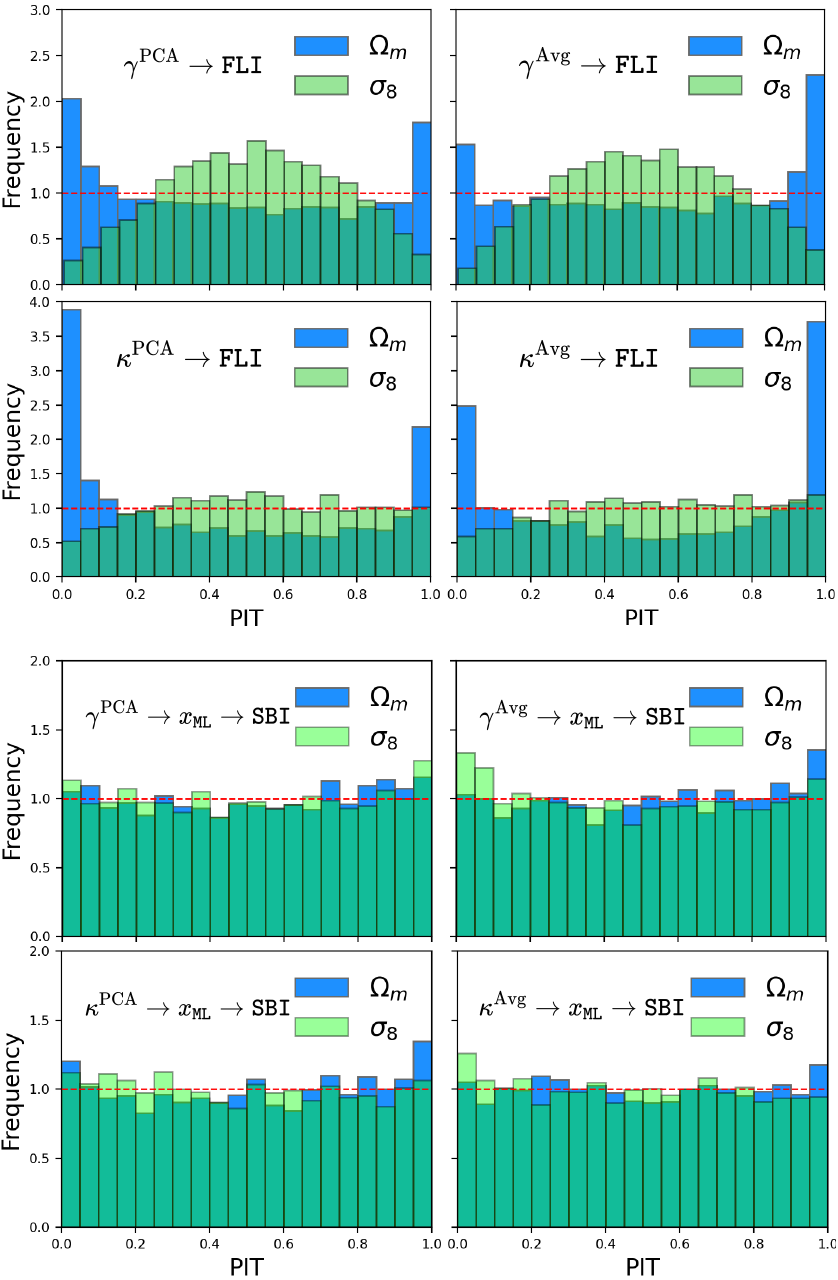}
\captionof{figure}{Comparison of PIT validation results between field-level inference (FLI) and ML-derived simulation-based inference (SBI), performed using four types of input data: (${\bm\gamma}^{\rm PCA}$, ${\bm\gamma}^{\rm Avg}$, ${\kappa}^{\rm PCA}$, ${\kappa}^{\rm Avg}$). Compared with FLI (top panels), the PIT distribution of SBI (bottom panels) is closer to a uniform distribution, indicating that the SBI posteriors are better calibrated and more consistent with the true parameters.}
\label{fig: PIT validation}
\end{center}

\begin{figure*}
\centering
\includegraphics[width=0.80\textwidth]{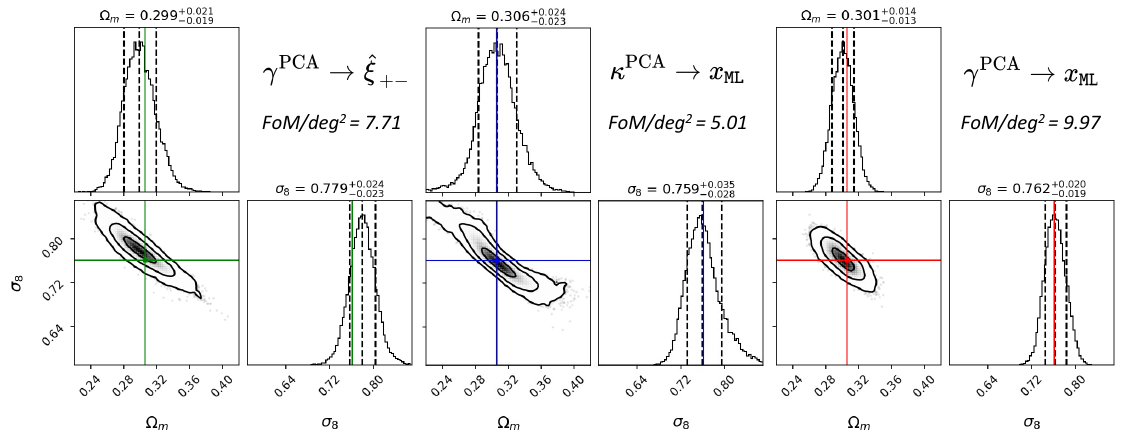}
\caption{\label{fig: SBI-MsmPCA: 2PCF-kappaML-gammaML}
Posteriors inferred via SBI using three types of summary statistics over the sky area of 655.4 ${\rm deg}^2$, with true cosmological parameters $\Omega^{\rm true}_{\rm m}=0.3062$ and $\sigma^{\rm true}_{8}=0.7615$. The left panel shows the posterior using the 2PCF ($\hat{\xi}_{+-}$) of the PCA-denoised shear field ($\gamma^{\rm PCA}$). The middle panel uses ML-derived features from the convergence field ($\kappa^{\rm PCA}$), reconstructed from $\gamma^{\rm PCA}$ via the \texttt{KS} method. The right panel uses ML-derived features directly extracted from $\gamma^{\rm PCA}$. Based on the values of ${\rm FoM}/{\rm deg}^{2}$, ML-derived features improve the constraining power by $29.3\%$ over the 2PCFs, and the shear-based method outperforms the kappa-based method by $99.0\%$.}
\end{figure*}

\subsection{Cosmological Outputs}
SBI offers a flexible and statistically rigorous framework for cosmological parameter estimation from complex data. In this study, we employ two distinct forms of summary statistics: (1) ML-derived feature representation ($x_{\texttt{ML}}$) and (2) the shear 2PCF ($\hat{\xi}_{+-}$). The constraining power on the target cosmological parameters is quantified by the Figure of Merit (FoM), normalized per square degree. 
\par 

In principle, given a fixed neural compressor, stochastic variations across independent SBI realizations should be negligible if the neural density estimator (NDE) is fully converged. In practice, however, factors such as finite training data, limited model capacity, and random initialization of the NDE can introduce small variations in the inferred posteriors.
\par
Each SBI run yields a distinct approximate posterior conditioned on its learned parameters. This reflects the stochastic nature of NDE, including random initialization and stochastic optimization during training, which introduces variability across realizations. To assess the impact of these effects, we perform 10 independent SBI realizations using a fixed trained FLI network as the neural compressor. We find that the inferred posteriors are broadly consistent across runs, indicating good stability of the SBI procedure. Based on this consistency, we report results from a representative NDE model that achieves strong constraining power, in order to illustrate the achievable performance of the method. The full set of inferred posteriors from the 10 SBI runs is presented in Appendix \ref{sec: appendix}.
\par 
To demonstrate the SBI performance in each case, we adopt the fiducial cosmology characterized by ($\Omega_{m}^{\rm true},\,\sigma_{8}^{\rm true})\,=\,(0.3062, \,0.7615)$. Figure~\ref{fig: SBI-MsmPCA: 2PCF-kappaML-gammaML} presents the posterior distributions inferred from three SBI schemes, all applied to PCA-denoised weak lensing fields. The left panel shows the results based on 2PCF summaries derived from the PCA-denoised shear field ($\gamma^{\rm PCA}$), whereas the middle and right panels correspond to ML-derived features extracted from the PCA-denoised convergence ($\kappa^{\rm PCA}$) and shear ($\gamma^{\rm PCA}$), respectively. A comparison between the left and right panels reveals that ML-derived features enhance the parameter constraining power by 29.3\% relative to the 2PCF-based summaries. In general, ML-powered compressors are expected to provide more informative summary statistics. However, the comparison between the left and middle panels shows that the shear-based 2PCF achieves stronger constraining power than the convergence-based ML-derived features. This is mainly due to the limited accuracy of convergence-field reconstruction using the Kaiser–Squires inversion. For our adopted setup, consisting of a $6.4^\circ \times 6.4^\circ$ field of view with a 6 arcmin resolution, the reconstructed convergence maps deviate noticeably from the true field, indicating non-negligible reconstruction errors arising from information loss and systematics in the shear-to-convergence inversion, including finite-field effects, boundary artifacts, and pixelization. In contrast, shear is the directly observed quantity and preserves the original lensing information more faithfully. Therefore, performing inference directly in the shear domain can better retain cosmological information and avoid reconstruction-induced biases. Consistently, the shear-based ML representation outperforms its convergence-based counterpart by 99.0\%, further highlighting the advantage of direct shear-field analysis for cosmological parameter inference.
\par
To further assess the impact of PCA denoising on shear-field-based posterior inference, we compare the results of 2PCF-based and ML-derived SBI obtained from $\gamma^{\rm Avg}$ and $\gamma^{\rm PCA}$, respectively (Figure~\ref{fig: SBI: MsmAvg vs MsmPCA}). As shown in the left panels, the conventional 2PCF-based SBI shows little improvement from PCA denoising, as the process primarily enhances non-Gaussian components that are not effectively captured by the 2PCF method. In contrast, the ML-derived SBI (right panels) benefits substantially from PCA denoising, achieving a 25.7\% increase in constraining power and demonstrating its effectiveness in strengthening parameter inference. Overall, compared with the traditional 2PCF method, the combination of PCA denoising and ML-based feature extraction further boosts the constraining power by 36.4\%, highlighting the advantage of integrating these techniques for improved cosmological inference. The summarizing of the cosmological constraints is listed in Table~\ref{tab: cosmological constraints}.

\begin{center}
\includegraphics[width=0.90\columnwidth]{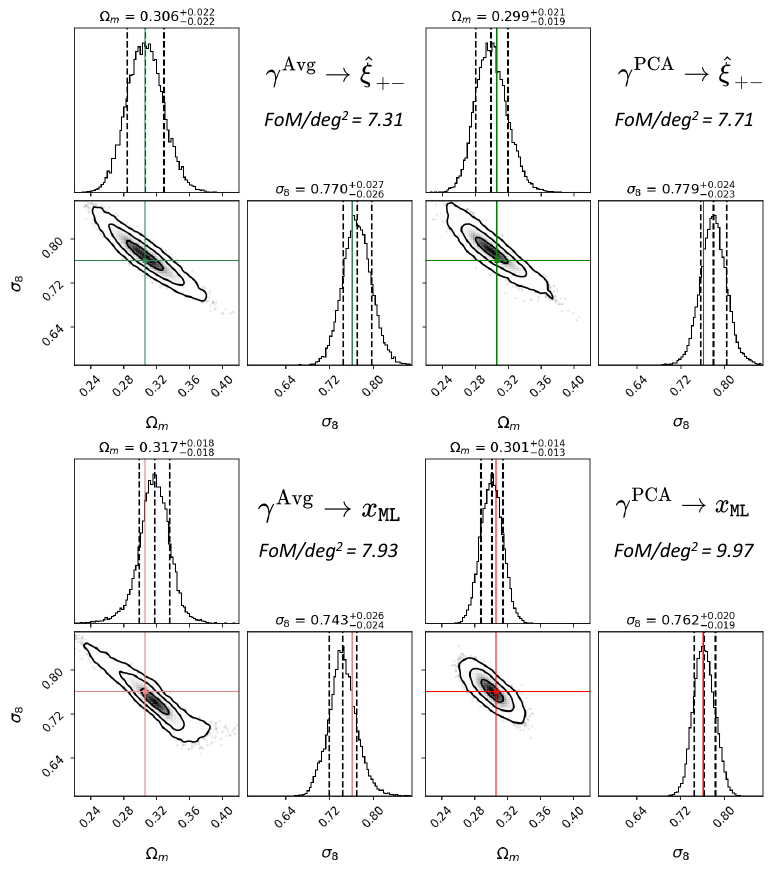}
\captionof{figure}{Comparison of SBI posteriors obtained from $\gamma^{\rm Avg}$ and $\gamma^{\rm PCA}$ over $655.4~{\rm deg}^2$, highlighting the impact of PCA denoising on two types of summary statistics: the 2PCFs ($\hat{\xi}_{+-}$; top panels) and ML-derived features ($x_{\texttt{ML}}$; bottom panels). Based on ${\rm FoM}/{\rm deg}^{2}$, PCA denoising improves the constraining power by $25.7\%$ when using ML-derived features.}
\label{fig: SBI: MsmAvg vs MsmPCA}
\end{center}

\begin{table*}
\renewcommand\arraystretch{1.4}
\centering
    \begin{tabular}{c|cc|c|cc|c}
    \hline\hline
    
    \multirow{2}{*}{Data} & 
    \multicolumn{2}{c|}{Denoising} & 
    \multirow{2}{*}{$\begin{array}{c}{\rm Summary} \\ {\rm Statistics}\end{array}$} & 
    \multirow{2}{*}{$\Omega_{m}$} & 
    \multirow{2}{*}{$\sigma_{8}$} &
    \multirow{2}{*}{${\rm FoM/deg^2}$} \\
    \cline{2-3}
                           & 
    $\mathcal{M}_{\rm sm}$ & 
    PCA                    & & & & \\ 
    \hline
    
    \multirow{2}{*}{$\gamma^{\rm Avg}$} & \multirow{2}{*}{\checkmark} & \multirow{2}{*}{\textbf{--}} & 2PCF & $0.306^{+0.022}_{-0.022}$ & $0.770^{+0.027}_{-0.026}$ & 7.31 \\  \cline{4-7}
      &  &  & ML-features & $0.317^{+0.018}_{-0.018}$ & $0.743^{+0.026}_{-0.024}$ & 7.93 \\
    \hline 
    \multirow{2}{*}{$\gamma^{\rm PCA}$} & \multirow{2}{*}{\checkmark} & \multirow{2}{*}{\checkmark} & 2PCF       & $0.302^{+0.021}_{-0.020}$ & $0.775^{+0.028}_{-0.028}$ & 7.71 \\  \cline{4-7}
      &  &  & ML-features & $0.301^{+0.014}_{-0.013}$ & $0.762^{+0.020}_{-0.019}$ & \textbf{9.97} \\
    \hline
    $\kappa^{\rm Avg}$                  & \checkmark & \textbf{--} & ML-features & $0.332^{+0.027}_{-0.025}$ & $0.717^{+0.028}_{-0.028}$ &  4.99 \\  
    \hline
    $\kappa^{\rm PCA}$                  & \checkmark & \checkmark  & ML-features & $0.306^{+0.024}_{-0.023}$ & $0.759^{+0.035}_{-0.028}$ & 5.01 \\  
    \hline\hline
    \end{tabular}
\caption{Cosmological constraints from SBI using a CSST-like WL survey covering 655.4 ${\rm deg}^2$, with the true parameters $(\Omega_{\rm m}^{\rm true},\sigma_8^{\rm true})$ $=$ $(0.3062, 0.7615)$.}
\label{tab: cosmological constraints}
\end{table*}

\section{Conclusions}
\label{sec: conclusion}
We have developed a ML-powered framework for shear-to-cosmology inference, motivated by the complementary strengths of simulation-based inference (SBI) and field-level inference (FLI). SBI offers a simple yet powerful approach for posterior estimation, capable of capturing complex, non-Gaussian, and even multi-modal parameter distributions. FLI networks can automatically extract informative and compact representations from cosmological fields. By integrating these two techniques, we establish an effective and robust shear-to-cosmology pipeline that harnesses the advantages of both SBI and FLI. This framework operates directly on the shear fields without requiring convergence reconstruction, and comprises three main components: (1) a shear denoising technique to mitigate the noise from intrinsic galaxy shapes, (2) a Transformer-based network for FLI, and (3) a SBI module.

\par
Theoretically, the convergence and shear fields contain equivalent cosmological information. In practice, however, the shear-to-convergence inversion amplifies noise and introduces filtering artifacts, resulting in non-negligible information loss. A shear-based analysis therefore offers a cleaner alternative. To assess its feasibility, we conducted preliminary experiments (not shown here) demonstrating that a Transformer-based U-Net can accurately reconstruct the convergence field from high–S/N shear data, indicating that the network is capable of learning the underlying reconstruction operator. Consequently, even with only shear inputs, the model can, in principle, recover information comparable to that in the true underlying convergence field. Consistently, SBI posteriors show that shear-based analysis nearly doubles the constraining power relative to the \texttt{KS}-convergence field, confirming that the FLI network does extract more cosmological information directly from shear. 
\par
In order to capture information encoded in cosmic evolution, the input to our framework is constructed from multi-redshift shear fields. In practice, increasing the number of redshift bins reduces the galaxy number density within each bin, thereby amplifying shape noise and making denoising an essential step in multi-redshift WL analyses. In this work, for shear-based analyses, we instead adopt PCA denoising along the redshift direction, which is a blind and training-free technique that suppresses noise while preserving the redshift evolution of the shear signal. To assess its impact, we evaluate the effect of PCA denoising using two types of summary statistics within the SBI framework: the 2PCFs and ML-derived features. For the 2PCF-based SBI, the improvement brought by PCA denoising is negligible. In contrast, when using ML-derived features, PCA denoising results in a 25.7\% increase in constraining power. This demonstrates that the PCA-denoised shear fields retain richer cosmological information, predominantly non-Gaussian components, which are effectively exploited by the ML-derived features but not by the traditional 2PCF analysis. 
\par
Several aspects of this work warrant further in-depth investigation. Weak lensing analyses have traditionally relied on the convergence field, reconstructed using a variety of methods, including the Kaiser–Squires algorithm~\citep{1993ApJ...404..441K}, Bayesian forward modeling~\citep{2016MNRAS.455.4452A}, Wiener filtering~\citep{2018MNRAS.479.2871J}, wavelet-based techniques~\citep{2006A&A...451.1139S,2025arXiv250903798G}, prior-free maximum-likelihood approaches~\citep{2024PhRvD.109l3530S,2025JCAP...07..038S}, and machine-learning methods~\citep{2020MNRAS.492.5023J,2024OJAp....7E..42S}. In this work, we adopt the standard \texttt{KS} algorithm for convergence reconstruction; however, a systematic comparison between shear-based inference and convergence-based results derived from alternative reconstruction techniques is still lacking and should be pursued in future studies.
\par
In addition, the impact of the denoising procedure has not yet been fully quantified. Our PCA-based approach, while simple and training-free, depends on several design choices (e.g., the number of subsets and the smoothing scale) and does not guarantee optimal separation between signal and noise. Moreover, as a linear decomposition method, PCA may not fully capture non-Gaussian or highly non-linear features in the shear field. These limitations suggest that the current denoising strategy can be further improved. For example, wavelet-based methods may provide a more natural separation of structures across different scales, while more advanced techniques such as ICA~\citep{HYVARINEN2000411} and GMCA~\citep{2007ITIP...16.2662B} offer promising alternatives for blind source separation. A systematic exploration of these approaches, as well as their impact on cosmological inference, remains an important direction for future work. More generally, combining map-level denoising with redshift-direction denoising could further enhance performance, although such a strategy would require substantially greater computational resources and is therefore left for future investigation.
\par 
Finally, due to computational limitations, both the angular resolution and the effective sky area of our input data are significantly lower than those expected for Stage-IV weak lensing surveys. In addition, several important observational systematics are not included in this preliminary study, such as PSF effects, intrinsic alignments, source clustering, and shear measurement biases. Addressing these limitations will require the construction of more realistic mock datasets, incorporating spectroscopic and photometric galaxy catalogs together with corresponding weak-lensing fields, based on high-resolution N-body simulations, partial-sky ray-tracing, and realistic image simulations.
\par
In conclusion, we present a unified machine learning framework that integrates FLI and SBI for direct shear-to-cosmology mapping. By combining PCA denoising, a transformer-driven feature extractor, and SBI posterior estimation, our approach enables accurate and efficient cosmological inference directly from shear fields. The results demonstrate that PCA denoising enhances the recovery of non-Gaussian information, while the Transformer-based network effectively captures cosmological signals beyond the reach of shear two-point analyses. These findings highlight the potential of ML-powered hybrid FLI–SBI frameworks to fully exploit the statistical power of weak lensing data, paving the way for more precise and information-rich cosmological parameter estimation in current and future surveys.

\begin{description}
    \item[] 
    \item[\textbf{Software}] TreeCorr~\citep{2004MNRAS.352..338J,2015ascl.soft08007J}, 
    DORIAN~\citep{2024MNRAS.533.3209F}, 
    PyTorch~\citep{2019arXiv191201703P}, 
    sbi~\citep{tejero-cantero2020sbi}, 
    HEALPix/healpy~\citep{2005ApJ...622..759G, Zonca2019},
    NumPy~\citep{van2011numpy,harris2020array},
    OpenMPI~\citep{gabriel2004open}, mpi4py~\citep{dalcin2021mpi4py},
    SciPy~\citep{virtanen2020scipy},
    Astropy~\citep{robitaille2013astropy,price2018astropy,price2022astropy},
    h5py~\citep{collette_python_hdf5_2014},
    Matplotlib~\citep{hunter2007matplotlib}, 
    corner~\citep{corner}
\end{description}


\begin{acknowledgements}
We acknowledge the support by National Key R\&D Program of China (2022YFF0503403), National Natural Science Foundation of China (12533008, 12503010, 12203084, 12573006, 12473097), China Manned Space Program (CMS-CSST-2021-(A02, A03, B01), CMS-CSST-2025-(A03, A05)), Postdoctoral Fellowship Program of CPSF (GZC20252089), Guangdong Basic and Applied Basic Research Foundation (2024A1515012309). JCD would like to thank “Tree New Bee” club for supporting friendly discussion environment.
\end{acknowledgements}

%


\bibliographystyle{aa}
\bibliography{reference}

@ARTICLE{2015RPPh...78h6901K,
       author = {{Kilbinger}, Martin},
        title = "{Cosmology with cosmic shear observations: a review}",
      journal = {Reports on Progress in Physics},
         year = 2015,
        month = jul,
       volume = {78},
       number = {8},
          eid = {086901},
        pages = {086901},
          doi = {10.1088/0034-4885/78/8/086901},
archivePrefix = {arXiv},
       eprint = {1411.0115},
 primaryClass = {astro-ph.CO},
       adsurl = {https://ui.adsabs.harvard.edu/abs/2015RPPh...78h6901K}
}

@BOOK{2022iglp.book.....M,
       author = {{Meneghetti}, Massimo},
        title = "{Introduction to Gravitational Lensing: With Python Examples}",
         year = 2022,
       adsurl = {https://ui.adsabs.harvard.edu/abs/2022iglp.book.....M}
}

@ARTICLE{2018ARA&A..56..393M,
       author = {{Mandelbaum}, Rachel},
        title = "{Weak Lensing for Precision Cosmology}",
      journal = {\araa},
         year = 2018,
        month = sep,
       volume = {56},
        pages = {393-433},
          doi = {10.1146/annurev-astro-081817-051928},
archivePrefix = {arXiv},
       eprint = {1710.03235},
 primaryClass = {astro-ph.CO},
       adsurl = {https://ui.adsabs.harvard.edu/abs/2018ARA&A..56..393M}
}

@ARTICLE{2009ApJ...695..652B,
       author = {{Bernstein}, Gary M.},
        title = "{Comprehensive Two-Point Analyses of Weak Gravitational Lensing Surveys}",
      journal = {\apj},
         year = 2009,
        month = apr,
       volume = {695},
       number = {1},
        pages = {652-665},
          doi = {10.1088/0004-637X/695/1/652},
archivePrefix = {arXiv},
       eprint = {0808.3400},
 primaryClass = {astro-ph},
       adsurl = {https://ui.adsabs.harvard.edu/abs/2009ApJ...695..652B}
}

@ARTICLE{2002A&A...396....1S,
       author = {{Schneider}, P. and {van Waerbeke}, L. and {Kilbinger}, M. and {Mellier}, Y.},
        title = "{Analysis of two-point statistics of cosmic shear. I. Estimators and covariances}",
      journal = {\aap},
         year = 2002,
        month = dec,
       volume = {396},
        pages = {1-19},
          doi = {10.1051/0004-6361:20021341},
archivePrefix = {arXiv},
       eprint = {astro-ph/0206182},
 primaryClass = {astro-ph},
       adsurl = {https://ui.adsabs.harvard.edu/abs/2002A&A...396....1S}
}

@ARTICLE{2209.04662,
  author = { {Kacprzak}, Tomasz and {Fluri}, Janis and {Schneider}, Aurel and {Refregier}, Alexandre and {Stadel}, Joachim},
  title = "{CosmoGridV1: a simulated wCDM theory prediction for map-level cosmological inference}",
  journal = {arXiv e-prints},
  year = 2022,
  month = sep,
  eid = {arXiv:2209.04662},
  pages = {arXiv:2209.04662},
  archivePrefix = {arXiv},
  eprint = {2209.04662},
  primaryClass = {astro-ph.CO}
  }

@article{PhysRevD.105.083518,
  title = {Full $w\mathrm{CDM}$ analysis of KiDS-1000 weak lensing maps using deep learning},
  author = { Fluri, Janis and Kacprzak, Tomasz and Lucchi, Aurelien and Schneider, Aurel and Refregier, Alexandre and Hofmann, Thomas},
  journal = {Phys. Rev. D},
  volume = {105},
  issue = {8},
  pages = {083518},
  numpages = {22},
  year = {2022},
  month = {Apr},
  publisher = {American Physical Society},
  doi = {10.1103/PhysRevD.105.083518},
  url = {https://link.aps.org/doi/10.1103/PhysRevD.105.083518}}

@ARTICLE{2024MNRAS.533.3209F,
       author = {{Ferlito}, Fulvio and {Davies}, Christopher T. and {Springel}, Volker and {Reinecke}, Martin and {Greco}, Alessandro and {Delgado}, Ana Maria and {White}, Simon D.~M. and {Hern{\'a}ndez-Aguayo}, C{\'e}sar and {Bose}, Sownak and {Hernquist}, Lars},
        title = "{Ray-tracing versus Born approximation in full-sky weak lensing simulations of the MillenniumTNG project}",
      journal = {\mnras},
         year = 2024,
        month = sep,
       volume = {533},
       number = {3},
        pages = {3209-3221},
          doi = {10.1093/mnras/stae2019},
archivePrefix = {arXiv},
       eprint = {2406.08540},
 primaryClass = {astro-ph.CO},
       adsurl = {https://ui.adsabs.harvard.edu/abs/2024MNRAS.533.3209F}
}

@ARTICLE{2019ApJ...883..203G,
       author = {{Gong}, Yan and {Liu}, Xiangkun and {Cao}, Ye and {Chen}, Xuelei and {Fan}, Zuhui and {Li}, Ran and {Li}, Xiao-Dong and {Li}, Zhigang and {Zhang}, Xin and {Zhan}, Hu},
        title = "{Cosmology from the Chinese Space Station Optical Survey (CSS-OS)}",
      journal = {\apj},
         year = 2019,
        month = oct,
       volume = {883},
       number = {2},
          eid = {203},
        pages = {203},
          doi = {10.3847/1538-4357/ab391e},
archivePrefix = {arXiv},
       eprint = {1901.04634},
 primaryClass = {astro-ph.CO},
       adsurl = {https://ui.adsabs.harvard.edu/abs/2019ApJ...883..203G}
}

@article{zhan2021wide,
  title={The wide-field multiband imaging and slitless spectroscopy survey to be carried out by the Survey Space Telescope of China Manned Space Program},
  author={Zhan, Hu},
  journal={Chinese Science Bulletin (Chinese Version)},
  volume={66},
  number={11},
  pages={1290--1298},
  year={2021},
  publisher={SCIENCE IN CHINA PRESS}
}

@ARTICLE{2018MNRAS.480.2178C,
       author = {{Cao}, Ye and {Gong}, Yan and {Meng}, Xian-Min and {Xu}, Cong K. and {Chen}, Xuelei and {Guo}, Qi and {Li}, Ran and {Liu}, Dezi and {Xue}, Yongquan and {Cao}, Li and {Fu}, Xiyang and {Zhang}, Xin and {Wang}, Shen and {Zhan}, Hu},
        title = "{Testing photometric redshift measurements with filter definition of the Chinese Space Station Optical Survey (CSS-OS)}",
      journal = {\mnras},
         year = 2018,
        month = oct,
       volume = {480},
       number = {2},
        pages = {2178-2190},
          doi = {10.1093/mnras/sty1980},
archivePrefix = {arXiv},
       eprint = {1706.09586},
 primaryClass = {astro-ph.IM},
       adsurl = {https://ui.adsabs.harvard.edu/abs/2018MNRAS.480.2178C}
}

@ARTICLE{2024SCPMA..6770413L,
       author = {{Liu}, Zhenjie and {Zhang}, Jun and {Li}, Hekun and {Shen}, Zhi and {Liu}, Cong},
        title = "{Quasi-2D weak lensing cosmological constraints using the PDF-SYM method}",
      journal = {\scpma},
         year = 2024,
        month = jul,
       volume = {67},
       number = {7},
          eid = {270413},
        pages = {270413},
          doi = {10.1007/s11433-024-2379-0},
archivePrefix = {arXiv},
       eprint = {2310.11209},
 primaryClass = {astro-ph.CO},
       adsurl = {https://ui.adsabs.harvard.edu/abs/2024SCPMA..6770413L}
}

@ARTICLE{2024MNRAS.527.5206Y,
       author = {{Yao}, Ji and {Shan}, Huanyuan and {Li}, Ran and {Xu}, Youhua and {Fan}, Dongwei and {Liu}, Dezi and {Zhang}, Pengjie and {Yu}, Yu and {Wei}, Chengliang and {Hu}, Bin and {Li}, Nan and {Fan}, Zuhui and {Xu}, Haojie and {Guo}, Wuzheng},
        title = "{CSST WL preparation I: forecast the impact from non-Gaussian covariances and requirements on systematics control}",
      journal = {\mnras},
         year = 2024,
        month = jan,
       volume = {527},
       number = {3},
        pages = {5206-5218},
          doi = {10.1093/mnras/stad3563},
archivePrefix = {arXiv},
       eprint = {2304.04489},
 primaryClass = {astro-ph.CO},
       adsurl = {https://ui.adsabs.harvard.edu/abs/2024MNRAS.527.5206Y}
}

@ARTICLE{2004MNRAS.352..338J,
       author = {{Jarvis}, M. and {Bernstein}, G. and {Jain}, B.},
        title = "{The skewness of the aperture mass statistic}",
      journal = {\mnras},
         year = 2004,
        month = jul,
       volume = {352},
       number = {1},
        pages = {338-352},
          doi = {10.1111/j.1365-2966.2004.07926.x},
archivePrefix = {arXiv},
       eprint = {astro-ph/0307393},
 primaryClass = {astro-ph},
       adsurl = {https://ui.adsabs.harvard.edu/abs/2004MNRAS.352..338J}
}

@software{2015ascl.soft08007J,
       author = {{Jarvis}, Mike},
        title = {TreeCorr: Two-point correlation functions},
 howpublished = {Astrophysics Source Code Library, record ascl:1508.007},
         year = 2015,
        month = aug,
          eid = {ascl:1508.007},
       adsurl = {https://ui.adsabs.harvard.edu/abs/2015ascl.soft08007J}
}

@ARTICLE{2020A&A...633A..69H,
       author = {{Hildebrandt}, H. and {K{\"o}hlinger}, F. and {van den Busch}, J.~L. and {Joachimi}, B. and {Heymans}, C. and {Kannawadi}, A. and {Wright}, A.~H. and {Asgari}, M. and {Blake}, C. and {Hoekstra}, H. and {Joudaki}, S. and {Kuijken}, K. and {Miller}, L. and {Morrison}, C.~B. and {Tr{\"o}ster}, T. and {Amon}, A. and {Archidiacono}, M. and {Brieden}, S. and {Choi}, A. and {de Jong}, J.~T.~A. and {Erben}, T. and {Giblin}, B. and {Mead}, A. and {Peacock}, J.~A. and {Radovich}, M. and {Schneider}, P. and {Sif{\'o}n}, C. and {Tewes}, M.},
        title = "{KiDS+VIKING-450: Cosmic shear tomography with optical and infrared data}",
      journal = {\aap},
         year = 2020,
        month = jan,
       volume = {633},
          eid = {A69},
        pages = {A69},
          doi = {10.1051/0004-6361/201834878},
archivePrefix = {arXiv},
       eprint = {1812.06076},
 primaryClass = {astro-ph.CO},
       adsurl = {https://ui.adsabs.harvard.edu/abs/2020A&A...633A..69H}
}

@ARTICLE{2020A&A...642A.158B,
       author = {{Blake}, Chris and {Amon}, Alexandra and {Asgari}, Marika and {Bilicki}, Maciej and {Dvornik}, Andrej and {Erben}, Thomas and {Giblin}, Benjamin and {Glazebrook}, Karl and {Heymans}, Catherine and {Hildebrandt}, Hendrik and {Joachimi}, Benjamin and {Joudaki}, Shahab and {Kannawadi}, Arun and {Kuijken}, Konrad and {Lidman}, Chris and {Parkinson}, David and {Shan}, HuanYuan and {Tr{\"o}ster}, Tilman and {van den Busch}, Jan Luca and {Wolf}, Christian and {Wright}, Angus H.},
        title = "{Testing gravity using galaxy-galaxy lensing and clustering amplitudes in KiDS-1000, BOSS, and 2dFLenS}",
      journal = {\aap},
         year = 2020,
        month = oct,
       volume = {642},
          eid = {A158},
        pages = {A158},
          doi = {10.1051/0004-6361/202038505},
archivePrefix = {arXiv},
       eprint = {2005.14351},
 primaryClass = {astro-ph.CO},
       adsurl = {https://ui.adsabs.harvard.edu/abs/2020A&A...642A.158B}
}

@ARTICLE{1993ApJ...404..441K,
       author = {{Kaiser}, Nick and {Squires}, Gordon},
        title = "{Mapping the Dark Matter with Weak Gravitational Lensing}",
      journal = {\apj},
         year = 1993,
        month = feb,
       volume = {404},
        pages = {441},
          doi = {10.1086/172297},
       adsurl = {https://ui.adsabs.harvard.edu/abs/1993ApJ...404..441K}
}

@ARTICLE{2010MNRAS.402.1049D,
       author = {{Dietrich}, J.~P. and {Hartlap}, J.},
        title = "{Cosmology with the shear-peak statistics}",
      journal = {\mnras},
         year = 2010,
        month = feb,
       volume = {402},
       number = {2},
        pages = {1049-1058},
          doi = {10.1111/j.1365-2966.2009.15948.x},
archivePrefix = {arXiv},
       eprint = {0906.3512},
 primaryClass = {astro-ph.CO},
       adsurl = {https://ui.adsabs.harvard.edu/abs/2010MNRAS.402.1049D}
}

@ARTICLE{2018MNRAS.474.1116S,
       author = {{Shan}, HuanYuan and {Liu}, Xiangkun and {Hildebrandt}, Hendrik and {Pan}, Chuzhong and {Martinet}, Nicolas and {Fan}, Zuhui and {Schneider}, Peter and {Asgari}, Marika and {Harnois-D{\'e}raps}, Joachim and {Hoekstra}, Henk and {Wright}, Angus and {Dietrich}, J{\"o}rg P. and {Erben}, Thomas and {Getman}, Fedor and {Grado}, Aniello and {Heymans}, Catherine and {Klaes}, Dominik and {Kuijken}, Konrad and {Merten}, Julian and {Puddu}, Emanuella and {Radovich}, Mario and {Wang}, Qiao},
        title = "{KiDS-450: cosmological constraints from weak lensing peak statistics - I. Inference from analytical prediction of high signal-to-noise ratio convergence peaks}",
      journal = {\mnras},
         year = 2018,
        month = feb,
       volume = {474},
       number = {1},
        pages = {1116-1134},
          doi = {10.1093/mnras/stx2837},
archivePrefix = {arXiv},
       eprint = {1709.07651},
 primaryClass = {astro-ph.CO},
       adsurl = {https://ui.adsabs.harvard.edu/abs/2018MNRAS.474.1116S}
}

@ARTICLE{2012PhRvD..85j3513K,
       author = {{Kratochvil}, Jan M. and {Lim}, Eugene A. and {Wang}, Sheng and {Haiman}, Zolt{\'a}n and {May}, Morgan and {Huffenberger}, Kevin},
        title = "{Probing cosmology with weak lensing Minkowski functionals}",
      journal = {\prd},
         year = 2012,
        month = may,
       volume = {85},
       number = {10},
          eid = {103513},
        pages = {103513},
          doi = {10.1103/PhysRevD.85.103513},
archivePrefix = {arXiv},
       eprint = {1109.6334},
 primaryClass = {astro-ph.CO},
       adsurl = {https://ui.adsabs.harvard.edu/abs/2012PhRvD..85j3513K}
}

@ARTICLE{2013PhRvD..88l3002P,
       author = {{Petri}, Andrea and {Haiman}, Zolt{\'a}n and {Hui}, Lam and {May}, Morgan and {Kratochvil}, Jan M.},
        title = "{Cosmology with Minkowski functionals and moments of the weak lensing convergence field}",
      journal = {\prd},
         year = 2013,
        month = dec,
       volume = {88},
       number = {12},
          eid = {123002},
        pages = {123002},
          doi = {10.1103/PhysRevD.88.123002},
archivePrefix = {arXiv},
       eprint = {1309.4460},
 primaryClass = {astro-ph.CO},
       adsurl = {https://ui.adsabs.harvard.edu/abs/2013PhRvD..88l3002P}
}

@ARTICLE{2019PhRvD.100f3514F,
       author = {{Fluri}, Janis and {Kacprzak}, Tomasz and {Lucchi}, Aurelien and {Refregier}, Alexandre and {Amara}, Adam and {Hofmann}, Thomas and {Schneider}, Aurel},
        title = "{Cosmological constraints with deep learning from KiDS-450 weak lensing maps}",
      journal = {\prd},
         year = 2019,
        month = sep,
       volume = {100},
       number = {6},
          eid = {063514},
        pages = {063514},
          doi = {10.1103/PhysRevD.100.063514},
archivePrefix = {arXiv},
       eprint = {1906.03156},
 primaryClass = {astro-ph.CO},
       adsurl = {https://ui.adsabs.harvard.edu/abs/2019PhRvD.100f3514F}
}

@ARTICLE{2007Natur.445..286M,
       author = {{Massey}, Richard and {Rhodes}, Jason and {Ellis}, Richard and {Scoville}, Nick and {Leauthaud}, Alexie and {Finoguenov}, Alexis and {Capak}, Peter and {Bacon}, David and {Aussel}, Herv{\'e} and {Kneib}, Jean-Paul and {Koekemoer}, Anton and {McCracken}, Henry and {Mobasher}, Bahram and {Pires}, Sandrine and {Refregier}, Alexandre and {Sasaki}, Shunji and {Starck}, Jean-Luc and {Taniguchi}, Yoshi and {Taylor}, Andy and {Taylor}, James},
        title = "{Dark matter maps reveal cosmic scaffolding}",
      journal = {\nat},
         year = 2007,
        month = jan,
       volume = {445},
       number = {7125},
        pages = {286-290},
          doi = {10.1038/nature05497},
archivePrefix = {arXiv},
       eprint = {astro-ph/0701594},
 primaryClass = {astro-ph},
       adsurl = {https://ui.adsabs.harvard.edu/abs/2007Natur.445..286M}
}

@ARTICLE{2013MNRAS.433.3373V,
       author = {{Van Waerbeke}, L. and {Benjamin}, J. and {Erben}, T. and {Heymans}, C. and {Hildebrandt}, H. and {Hoekstra}, H. and {Kitching}, T.~D. and {Mellier}, Y. and {Miller}, L. and {Coupon}, J. and {Harnois-D{\'e}raps}, J. and {Fu}, L. and {Hudson}, M. and {Kilbinger}, M. and {Kuijken}, K. and {Rowe}, B. and {Schrabback}, T. and {Semboloni}, E. and {Vafaei}, S. and {van Uitert}, E. and {Velander}, M.},
        title = "{CFHTLenS: mapping the large-scale structure with gravitational lensing}",
      journal = {\mnras},
         year = 2013,
        month = aug,
       volume = {433},
       number = {4},
        pages = {3373-3388},
          doi = {10.1093/mnras/stt971},
archivePrefix = {arXiv},
       eprint = {1303.1806},
 primaryClass = {astro-ph.CO},
       adsurl = {https://ui.adsabs.harvard.edu/abs/2013MNRAS.433.3373V}
}

@ARTICLE{2018PASJ...70S..26O,
       author = {{Oguri}, Masamune and {Miyazaki}, Satoshi and {Hikage}, Chiaki and {Mandelbaum}, Rachel and {Utsumi}, Yousuke and {Miyatake}, Hironao and {Takada}, Masahiro and {Armstrong}, Robert and {Bosch}, James and {Komiyama}, Yutaka and {Leauthaud}, Alexie and {More}, Surhud and {Nishizawa}, Atsushi J. and {Okabe}, Nobuhiro and {Tanaka}, Masayuki},
        title = "{Two- and three-dimensional wide-field weak lensing mass maps from the Hyper Suprime-Cam Subaru Strategic Program S16A data}",
      journal = {\pasj},
         year = 2018,
        month = jan,
       volume = {70},
          eid = {S26},
        pages = {S26},
          doi = {10.1093/pasj/psx070},
archivePrefix = {arXiv},
       eprint = {1705.06792},
 primaryClass = {astro-ph.CO},
       adsurl = {https://ui.adsabs.harvard.edu/abs/2018PASJ...70S..26O}
}

@ARTICLE{2014MNRAS.442.2534S,
       author = {{Shan}, Huan Yuan and {Kneib}, Jean-Paul and {Comparat}, Johan and {Jullo}, Eric and {Charbonnier}, Ald{\'e}e and {Erben}, Thomas and {Makler}, Martin and {Moraes}, Bruno and {Van Waerbeke}, Ludovic and {Courbin}, Fr{\'e}d{\'e}ric and {Meylan}, Georges and {Tao}, Charling and {Taylor}, James E.},
        title = "{Weak lensing mass map and peak statistics in Canada-France-Hawaii Telescope Stripe 82 survey}",
      journal = {\mnras},
         year = 2014,
        month = aug,
       volume = {442},
       number = {3},
        pages = {2534-2542},
          doi = {10.1093/mnras/stu1040},
archivePrefix = {arXiv},
       eprint = {1311.1319},
 primaryClass = {astro-ph.CO},
       adsurl = {https://ui.adsabs.harvard.edu/abs/2014MNRAS.442.2534S}
}

@ARTICLE{2005ApJ...635...60T,
       author = {{Tang}, J.~Y. and {Fan}, Z.~H.},
        title = "{Effects of the Complex Mass Distribution of Dark Matter Halos on Weak-Lensing Cluster Surveys}",
      journal = {\apj},
         year = 2005,
        month = dec,
       volume = {635},
       number = {1},
        pages = {60-72},
          doi = {10.1086/497285},
archivePrefix = {arXiv},
       eprint = {astro-ph/0508508},
 primaryClass = {astro-ph},
       adsurl = {https://ui.adsabs.harvard.edu/abs/2005ApJ...635...60T}
}

@ARTICLE{2012ApJ...748...56S,
       author = {{Shan}, HuanYuan and {Kneib}, Jean-Paul and {Tao}, Charling and {Fan}, Zuhui and {Jauzac}, Mathilde and {Limousin}, Marceau and {Massey}, Richard and {Rhodes}, Jason and {Thanjavur}, Karun and {McCracken}, Henry J.},
        title = "{Weak Lensing Measurement of Galaxy Clusters in the CFHTLS-Wide Survey}",
      journal = {\apj},
         year = 2012,
        month = mar,
       volume = {748},
       number = {1},
          eid = {56},
        pages = {56},
          doi = {10.1088/0004-637X/748/1/56},
archivePrefix = {arXiv},
       eprint = {1108.1981},
 primaryClass = {astro-ph.CO},
       adsurl = {https://ui.adsabs.harvard.edu/abs/2012ApJ...748...56S}
}

@ARTICLE{2004ApJ...606...67H,
       author = {{Hoekstra}, Henk and {Yee}, H.~K.~C. and {Gladders}, Michael D.},
        title = "{Properties of Galaxy Dark Matter Halos from Weak Lensing}",
      journal = {\apj},
         year = 2004,
        month = may,
       volume = {606},
       number = {1},
        pages = {67-77},
          doi = {10.1086/382726},
archivePrefix = {arXiv},
       eprint = {astro-ph/0306515},
 primaryClass = {astro-ph},
       adsurl = {https://ui.adsabs.harvard.edu/abs/2004ApJ...606...67H}
}

@ARTICLE{1999ApJ...522L..21H,
       author = {{Hu}, Wayne},
        title = "{Power Spectrum Tomography with Weak Lensing}",
      journal = {\apjl},
         year = 1999,
        month = sep,
       volume = {522},
       number = {1},
        pages = {L21-L24},
          doi = {10.1086/312210},
archivePrefix = {arXiv},
       eprint = {astro-ph/9904153},
 primaryClass = {astro-ph},
       adsurl = {https://ui.adsabs.harvard.edu/abs/1999ApJ...522L..21H}
}

@ARTICLE{2013MNRAS.431.1547B,
       author = {{Benjamin}, Jonathan and {Van Waerbeke}, Ludovic and {Heymans}, Catherine and {Kilbinger}, Martin and {Erben}, Thomas and {Hildebrandt}, Hendrik and {Hoekstra}, Henk and {Kitching}, Thomas D. and {Mellier}, Yannick and {Miller}, Lance and {Rowe}, Barnaby and {Schrabback}, Tim and {Simpson}, Fergus and {Coupon}, Jean and {Fu}, Liping and {Harnois-D{\'e}raps}, Joachim and {Hudson}, Michael J. and {Kuijken}, Konrad and {Semboloni}, Elisabetta and {Vafaei}, Sanaz and {Velander}, Malin},
        title = "{CFHTLenS tomographic weak lensing: quantifying accurate redshift distributions}",
      journal = {\mnras},
         year = 2013,
        month = may,
       volume = {431},
       number = {2},
        pages = {1547-1564},
          doi = {10.1093/mnras/stt276},
archivePrefix = {arXiv},
       eprint = {1212.3327},
 primaryClass = {astro-ph.CO},
       adsurl = {https://ui.adsabs.harvard.edu/abs/2013MNRAS.431.1547B}
}

@ARTICLE{2010A&A...516A..63S,
       author = {{Schrabback}, T. and {Hartlap}, J. and {Joachimi}, B. and {Kilbinger}, M. and {Simon}, P. and {Benabed}, K. and {Brada{\v{c}}}, M. and {Eifler}, T. and {Erben}, T. and {Fassnacht}, C.~D. and {High}, F. William and {Hilbert}, S. and {Hildebrandt}, H. and {Hoekstra}, H. and {Kuijken}, K. and {Marshall}, P.~J. and {Mellier}, Y. and {Morganson}, E. and {Schneider}, P. and {Semboloni}, E. and {van Waerbeke}, L. and {Velander}, M.},
        title = "{Evidence of the accelerated expansion of the Universe from weak lensing tomography with COSMOS}",
      journal = {\aap},
         year = 2010,
        month = jun,
       volume = {516},
          eid = {A63},
        pages = {A63},
          doi = {10.1051/0004-6361/200913577},
archivePrefix = {arXiv},
       eprint = {0911.0053},
 primaryClass = {astro-ph.CO},
       adsurl = {https://ui.adsabs.harvard.edu/abs/2010A&A...516A..63S}
}

@ARTICLE{2020MNRAS.499.5902C,
       author = {{Cheng}, Sihao and {Ting}, Yuan-Sen and {M{\'e}nard}, Brice and {Bruna}, Joan},
        title = "{A new approach to observational cosmology using the scattering transform}",
      journal = {\mnras},
         year = 2020,
        month = dec,
       volume = {499},
       number = {4},
        pages = {5902-5914},
          doi = {10.1093/mnras/staa3165},
archivePrefix = {arXiv},
       eprint = {2006.08561},
 primaryClass = {astro-ph.CO},
       adsurl = {https://ui.adsabs.harvard.edu/abs/2020MNRAS.499.5902C}
}

@ARTICLE{2025JCAP...01..006C,
       author = {{Cheng}, Sihao and {Marques}, Gabriela A. and {Grand{\'o}n}, Daniela and {Thiele}, Leander and {Shirasaki}, Masato and {M{\'e}nard}, Brice and {Liu}, Jia},
        title = "{Cosmological constraints from weak lensing scattering transform using HSC Y1 data}",
      journal = {\jcap},
         year = 2025,
        month = jan,
       volume = {2025},
       number = {1},
          eid = {006},
        pages = {006},
          doi = {10.1088/1475-7516/2025/01/006},
archivePrefix = {arXiv},
       eprint = {2404.16085},
 primaryClass = {astro-ph.CO},
       adsurl = {https://ui.adsabs.harvard.edu/abs/2025JCAP...01..006C}
}

@ARTICLE{2020MNRAS.492.5023J,
       author = {{Jeffrey}, Niall and {Lanusse}, Fran{\c{c}}ois and {Lahav}, Ofer and {Starck}, Jean-Luc},
        title = "{Deep learning dark matter map reconstructions from DES SV weak lensing data}",
      journal = {\mnras},
         year = 2020,
        month = mar,
       volume = {492},
       number = {4},
        pages = {5023-5029},
          doi = {10.1093/mnras/staa127},
archivePrefix = {arXiv},
       eprint = {1908.00543},
 primaryClass = {astro-ph.CO},
       adsurl = {https://ui.adsabs.harvard.edu/abs/2020MNRAS.492.5023J}
}

@ARTICLE{2024A&A...683A.209Z,
       author = {{Zhang}, Zekang and {Shan}, Huanyuan and {Li}, Nan and {Wei}, Chengliang and {Yao}, Ji and {Ban}, Zhang and {Fang}, Yuedong and {Guo}, Qi and {Liu}, Dezi and {Li}, Guoliang and {Lin}, Lin and {Li}, Ming and {Li}, Ran and {Li}, Xiaobo and {Luo}, Yu and {Meng}, Xianmin and {Nie}, Jundan and {Qi}, Zhaoxiang and {Qiu}, Yisheng and {Shao}, Li and {Tian}, Hao and {Wang}, Lei and {Wang}, Wei and {Xian}, Jingtian and {Xu}, Youhua and {Zhang}, Tianmeng and {Zhang}, Xin and {Zhou}, Zhimin},
        title = "{FORKLENS: Accurate weak-lensing shear measurement with deep learning}",
      journal = {\aap},
         year = 2024,
        month = mar,
       volume = {683},
          eid = {A209},
        pages = {A209},
          doi = {10.1051/0004-6361/202345903},
archivePrefix = {arXiv},
       eprint = {2301.02986},
 primaryClass = {astro-ph.CO},
       adsurl = {https://ui.adsabs.harvard.edu/abs/2024A&A...683A.209Z}
}

@ARTICLE{2025A&A...699A.174S,
       author = {{Su}, Chen and {Shan}, Huanyuan and {Zhao}, Cheng and {Xu}, Wenshuo and {Zhang}, Jiajun},
        title = "{Cosmological constraints with void lensing: I. Simulation-based inference framework}",
      journal = {\aap},
         year = 2025,
        month = jul,
       volume = {699},
          eid = {A174},
        pages = {A174},
          doi = {10.1051/0004-6361/202555237},
archivePrefix = {arXiv},
       eprint = {2504.15149},
 primaryClass = {astro-ph.CO},
       adsurl = {https://ui.adsabs.harvard.edu/abs/2025A&A...699A.174S}
}

@article{tejero-cantero2020sbi,
  doi = {10.21105/joss.02505},
  url = {https://doi.org/10.21105/joss.02505},
  year = {2020},
  publisher = {The Open Journal},
  volume = {5},
  number = {52},
  pages = {2505},
  author = {Alvaro Tejero-Cantero and Jan Boelts and Michael Deistler and Jan-Matthis Lueckmann and Conor Durkan and Pedro J. Gonçalves and David S. Greenberg and Jakob H. Macke},
  title = {sbi: A toolkit for simulation-based inference},
  journal = {\joss}
}

@article{boelts2024sbi,
  title={sbi reloaded: a toolkit for simulation-based inference workflows},
  author={Boelts, Jan and Deistler, Michael and Gloeckler, Manuel and Tejero-Cantero, {\'A}lvaro and Lueckmann, Jan-Matthis and Moss, Guy and Steinbach, Peter and Moreau, Thomas and Muratore, Fabio and Linhart, Julia and others},
  journal={arXiv preprint arXiv:2411.17337},
  year={2024}
}

@article{leclercq2025field,
  title={Field-level inference in cosmology},
  author={Leclercq, Florent},
  journal={arXiv preprint arXiv:2509.13435},
  year={2025}
}

@ARTICLE{2025NatAs...9..608L,
       author = {{Liu}, Tie and {Quan}, Yuhui and {Su}, Yingna and {Guo}, Yang and {Liu}, Shu and {Ji}, Haisheng and {Hao}, Qi and {Gao}, Yulong and {Liu}, Yuxia and {Wang}, Yikang and {Sun}, Wenqing and {Ding}, Mingde},
        title = "{Astronomical image denoising by self-supervised deep learning and restoration processes}",
      journal = {\natA},
         year = 2025,
        month = apr,
       volume = {9},
        pages = {608-615},
          doi = {10.1038/s41550-025-02484-z},
archivePrefix = {arXiv},
       eprint = {2502.16807},
 primaryClass = {astro-ph.IM},
       adsurl = {https://ui.adsabs.harvard.edu/abs/2025NatAs...9..608L}
}

@ARTICLE{2017MNRAS.467L.110S,
       author = {{Schawinski}, Kevin and {Zhang}, Ce and {Zhang}, Hantian and {Fowler}, Lucas and {Santhanam}, Gokula Krishnan},
        title = "{Generative adversarial networks recover features in astrophysical images of galaxies beyond the deconvolution limit}",
      journal = {\mnras},
         year = 2017,
        month = may,
       volume = {467},
       number = {1},
        pages = {L110-L114},
          doi = {10.1093/mnrasl/slx008},
archivePrefix = {arXiv},
       eprint = {1702.00403},
 primaryClass = {astro-ph.IM},
       adsurl = {https://ui.adsabs.harvard.edu/abs/2017MNRAS.467L.110S}
}

@ARTICLE{2015MNRAS.450.1441D,
       author = {{Dieleman}, Sander and {Willett}, Kyle W. and {Dambre}, Joni},
        title = "{Rotation-invariant convolutional neural networks for galaxy morphology prediction}",
      journal = {\mnras},
         year = 2015,
        month = jun,
       volume = {450},
       number = {2},
        pages = {1441-1459},
          doi = {10.1093/mnras/stv632},
archivePrefix = {arXiv},
       eprint = {1503.07077},
 primaryClass = {astro-ph.IM},
       adsurl = {https://ui.adsabs.harvard.edu/abs/2015MNRAS.450.1441D}
}

@ARTICLE{2017MNRAS.472.1129P,
       author = {{Petrillo}, C.~E. and {Tortora}, C. and {Chatterjee}, S. and {Vernardos}, G. and {Koopmans}, L.~V.~E. and {Verdoes Kleijn}, G. and {Napolitano}, N.~R. and {Covone}, G. and {Schneider}, P. and {Grado}, A. and {McFarland}, J.},
        title = "{Finding strong gravitational lenses in the Kilo Degree Survey with Convolutional Neural Networks}",
      journal = {\mnras},
         year = 2017,
        month = nov,
       volume = {472},
       number = {1},
        pages = {1129-1150},
          doi = {10.1093/mnras/stx2052},
archivePrefix = {arXiv},
       eprint = {1702.07675},
 primaryClass = {astro-ph.GA},
       adsurl = {https://ui.adsabs.harvard.edu/abs/2017MNRAS.472.1129P}
}

@ARTICLE{2025RAA....25l5003W,
       author = {{Wu}, You and {Li}, Nan and {Shan}, Huan-Yuan and {Wei}, Peng and {Wei}, Cheng-Liang and {Nie}, Lin and {Ren}, Juan-Juan and {Ban}, Zhang and {Li}, Xiao-Bo and {Yang}, Xun and {Jiang}, Yu-Xi and {Ma}, Hong-Cai and {Wang}, Wei and {Liu}, Chao},
        title = "{Improving PSF Reconstruction for CSST: A Combined Approach with Deep Learning Source Selection and Empirical Correction}",
      journal = {\raa},
         year = 2025,
        month = dec,
       volume = {25},
       number = {12},
          eid = {125003},
        pages = {125003},
          doi = {10.1088/1674-4527/ae088a},
       adsurl = {https://ui.adsabs.harvard.edu/abs/2025RAA....25l5003W}
}

@ARTICLE{2025ApJS..277...12Z,
       author = {{Zhong}, Fucheng and {Luo}, Ruibiao and {Napolitano}, Nicola R. and {Tortora}, Crescenzo and {Li}, Rui and {Zhu}, Xincheng and {Busillo}, Valerio and {Koopmans}, L.~V.~E. and {Longo}, Giuseppe},
        title = "{Galaxy{\textendash}Galaxy Strong Lensing with U-Net (GGSL-UNet). I. Extracting Two-dimensional Information from Multiband Images in Ground and Space Observations}",
      journal = {\apjs},
         year = 2025,
        month = mar,
       volume = {277},
       number = {1},
          eid = {12},
        pages = {12},
          doi = {10.3847/1538-4365/ada609},
archivePrefix = {arXiv},
       eprint = {2410.02936},
 primaryClass = {astro-ph.GA},
       adsurl = {https://ui.adsabs.harvard.edu/abs/2025ApJS..277...12Z}
}

@ARTICLE{2024PhRvD.110f3531M,
       author = {{Min}, Zhiwei and {Xiao}, Xu and {Ding}, Jiacheng and {Xiao}, Liang and {Jiang}, Jie and {Wu}, Donglin and {Lin}, Qiufan and {Wang}, Yang and {Liu}, Shuai and {Chen}, Zhixin and {Li}, Xiangru and {Zhang}, Jinqu and {Zhang}, Le and {Li}, Xiao-Dong},
        title = "{Deep learning for cosmological parameter inference from a dark matter halo density field}",
      journal = {\prd},
         year = 2024,
        month = sep,
       volume = {110},
       number = {6},
          eid = {063531},
        pages = {063531},
          doi = {10.1103/PhysRevD.110.063531},
archivePrefix = {arXiv},
       eprint = {2404.09483},
 primaryClass = {astro-ph.CO},
       adsurl = {https://ui.adsabs.harvard.edu/abs/2024PhRvD.110f3531M}
}

@ARTICLE{2024MNRAS.527.1244S,
       author = {{Stopyra}, Stephen and {Peiris}, Hiranya V. and {Pontzen}, Andrew and {Jasche}, Jens and {Lavaux}, Guilhem},
        title = "{Towards accurate field-level inference of massive cosmic structures}",
      journal = {\mnras},
         year = 2024,
        month = jan,
       volume = {527},
       number = {1},
        pages = {1244-1256},
          doi = {10.1093/mnras/stad3170},
archivePrefix = {arXiv},
       eprint = {2304.09193},
 primaryClass = {astro-ph.CO},
       adsurl = {https://ui.adsabs.harvard.edu/abs/2024MNRAS.527.1244S}
}

@ARTICLE{2021MNRAS.506L..85L,
       author = {{Leclercq}, Florent and {Heavens}, Alan},
        title = "{On the accuracy and precision of correlation functions and field-level inference in cosmology}",
      journal = {\mnras},
         year = 2021,
        month = sep,
       volume = {506},
       number = {1},
        pages = {L85-L90},
          doi = {10.1093/mnrasl/slab081},
archivePrefix = {arXiv},
       eprint = {2103.04158},
 primaryClass = {astro-ph.CO},
       adsurl = {https://ui.adsabs.harvard.edu/abs/2021MNRAS.506L..85L}
}

@ARTICLE{2020PhRvD.102j3509H,
       author = {{Hort{\'u}a}, H{\'e}ctor J. and {Volpi}, Riccardo and {Marinelli}, Dimitri and {Malag{\`o}}, Luigi},
        title = "{Parameter estimation for the cosmic microwave background with Bayesian neural networks}",
      journal = {\prd},
         year = 2020,
        month = nov,
       volume = {102},
       number = {10},
          eid = {103509},
        pages = {103509},
          doi = {10.1103/PhysRevD.102.103509},
archivePrefix = {arXiv},
       eprint = {1911.08508},
 primaryClass = {astro-ph.IM},
       adsurl = {https://ui.adsabs.harvard.edu/abs/2020PhRvD.102j3509H}
}

@ARTICLE{2020PNAS..11730055C,
       author = {{Cranmer}, Kyle and {Brehmer}, Johann and {Louppe}, Gilles},
        title = "{The frontier of simulation-based inference}",
      journal = {\pnas},
         year = 2020,
        month = dec,
       volume = {117},
       number = {48},
        pages = {30055-30062},
          doi = {10.1073/pnas.1912789117},
archivePrefix = {arXiv},
       eprint = {1911.01429},
 primaryClass = {stat.ML},
       adsurl = {https://ui.adsabs.harvard.edu/abs/2020PNAS..11730055C}
}

@ARTICLE{2023MLS&T...4aLT01L,
       author = {{Lemos}, Pablo and {Cranmer}, Miles and {Abidi}, Muntazir and {Hahn}, ChangHoon and {Eickenberg}, Michael and {Massara}, Elena and {Yallup}, David and {Ho}, Shirley},
        title = "{Robust simulation-based inference in cosmology with Bayesian neural networks}",
      journal = {\MLst},
         year = 2023,
        month = mar,
       volume = {4},
       number = {1},
          eid = {01LT01},
        pages = {01LT01},
          doi = {10.1088/2632-2153/acbb53},
archivePrefix = {arXiv},
       eprint = {2207.08435},
 primaryClass = {astro-ph.CO},
       adsurl = {https://ui.adsabs.harvard.edu/abs/2023MLS&T...4aLT01L}
}

@ARTICLE{2025A&A...699A.327Z,
       author = {{Zeghal}, Justine and {Lanzieri}, Denise and {Lanusse}, Fran{\c{c}}ois and {Boucaud}, Alexandre and {Louppe}, Gilles and {Aubourg}, Eric and {Bayer}, Adrian E. and {LSST Dark Energy Science Collaboration}},
        title = "{Simulation-based inference benchmark for weak lensing cosmology}",
      journal = {\aap},
         year = 2025,
        month = jul,
       volume = {699},
          eid = {A327},
        pages = {A327},
          doi = {10.1051/0004-6361/202452410},
archivePrefix = {arXiv},
       eprint = {2409.17975},
 primaryClass = {astro-ph.CO},
       adsurl = {https://ui.adsabs.harvard.edu/abs/2025A&A...699A.327Z}
}

@ARTICLE{2023MNRAS.521.2050L,
       author = {{Lu}, Tianhuan and {Haiman}, Zolt{\'a}n and {Li}, Xiangchong},
        title = "{Cosmological constraints from HSC survey first-year data using deep learning}",
      journal = {\mnras},
         year = 2023,
        month = may,
       volume = {521},
       number = {2},
        pages = {2050-2066},
          doi = {10.1093/mnras/stad686},
archivePrefix = {arXiv},
       eprint = {2301.01354},
 primaryClass = {astro-ph.CO},
       adsurl = {https://ui.adsabs.harvard.edu/abs/2023MNRAS.521.2050L}
}

@ARTICLE{2016PhRvD..94d3533L,
       author = {{Liu}, Jia and {Haiman}, Zolt{\'a}n},
        title = "{Origin of weak lensing convergence peaks}",
      journal = {\prd},
         year = 2016,
        month = aug,
       volume = {94},
       number = {4},
          eid = {043533},
        pages = {043533},
          doi = {10.1103/PhysRevD.94.043533},
archivePrefix = {arXiv},
       eprint = {1606.01318},
 primaryClass = {astro-ph.CO},
       adsurl = {https://ui.adsabs.harvard.edu/abs/2016PhRvD..94d3533L}
}

@ARTICLE{2015MNRAS.447..400A,
       author = {{Alonso}, David and {Bull}, Philip and {Ferreira}, Pedro G. and {Santos}, M{\'a}rio G.},
        title = "{Blind foreground subtraction for intensity mapping experiments}",
      journal = {\mnras},
         year = 2015,
        month = feb,
       volume = {447},
       number = {1},
        pages = {400-416},
          doi = {10.1093/mnras/stu2474},
archivePrefix = {arXiv},
       eprint = {1409.8667},
 primaryClass = {astro-ph.CO},
       adsurl = {https://ui.adsabs.harvard.edu/abs/2015MNRAS.447..400A}
}

@ARTICLE{2021Univ....7..213P,
       author = {{Padilla}, Luis E. and {Tellez}, Luis O. and {Escamilla}, Luis A. and {Vazquez}, Jose Alberto},
        title = "{Cosmological Parameter Inference with Bayesian Statistics}",
      journal = {Universe},
         year = 2021,
        month = jun,
       volume = {7},
       number = {7},
          eid = {213},
        pages = {213},
          doi = {10.3390/universe7070213},
       adsurl = {https://ui.adsabs.harvard.edu/abs/2021Univ....7..213P}
}

@ARTICLE{2022PhRvD.106f3536T,
       author = {{Taylor}, Peter L. and {Markovi{\v{c}}}, Katarina},
        title = "{Covariance of photometric and spectroscopic two-point statistics: Implications for cosmological parameter inference}",
      journal = {\prd},
         year = 2022,
        month = sep,
       volume = {106},
       number = {6},
          eid = {063536},
        pages = {063536},
          doi = {10.1103/PhysRevD.106.063536},
archivePrefix = {arXiv},
       eprint = {2205.14167},
 primaryClass = {astro-ph.CO},
       adsurl = {https://ui.adsabs.harvard.edu/abs/2022PhRvD.106f3536T}
}

@ARTICLE{2020A&A...641A...6P,
       author = {{Planck Collaboration} and {Aghanim}, N. and {Akrami}, Y. and {Ashdown}, M. and {Aumont}, J. and {Baccigalupi}, C. and {Ballardini}, M. and {Banday}, A.~J. and {Barreiro}, R.~B. and {Bartolo}, N. and {Basak}, S. and {Battye}, R. and {Benabed}, K. and {Bernard}, J.-P. and {Bersanelli}, M. and {Bielewicz}, P. and {Bock}, J.~J. and {Bond}, J.~R. and {Borrill}, J. and {Bouchet}, F.~R. and {Boulanger}, F. and {Bucher}, M. and {Burigana}, C. and {Butler}, R.~C. and {Calabrese}, E. and {Cardoso}, J.-F. and {Carron}, J. and {Challinor}, A. and {Chiang}, H.~C. and {Chluba}, J. and {Colombo}, L.~P.~L. and {Combet}, C. and {Contreras}, D. and {Crill}, B.~P. and {Cuttaia}, F. and {de Bernardis}, P. and {de Zotti}, G. and {Delabrouille}, J. and {Delouis}, J.-M. and {Di Valentino}, E. and {Diego}, J.~M. and {Dor{\'e}}, O. and {Douspis}, M. and {Ducout}, A. and {Dupac}, X. and {Dusini}, S. and {Efstathiou}, G. and {Elsner}, F. and {En{\ss}lin}, T.~A. and {Eriksen}, H.~K. and {Fantaye}, Y. and {Farhang}, M. and {Fergusson}, J. and {Fernandez-Cobos}, R. and {Finelli}, F. and {Forastieri}, F. and {Frailis}, M. and {Fraisse}, A.~A. and {Franceschi}, E. and {Frolov}, A. and {Galeotta}, S. and {Galli}, S. and {Ganga}, K. and {G{\'e}nova-Santos}, R.~T. and {Gerbino}, M. and {Ghosh}, T. and {Gonz{\'a}lez-Nuevo}, J. and {G{\'o}rski}, K.~M. and {Gratton}, S. and {Gruppuso}, A. and {Gudmundsson}, J.~E. and {Hamann}, J. and {Handley}, W. and {Hansen}, F.~K. and {Herranz}, D. and {Hildebrandt}, S.~R. and {Hivon}, E. and {Huang}, Z. and {Jaffe}, A.~H. and {Jones}, W.~C. and {Karakci}, A. and {Keih{\"a}nen}, E. and {Keskitalo}, R. and {Kiiveri}, K. and {Kim}, J. and {Kisner}, T.~S. and {Knox}, L. and {Krachmalnicoff}, N. and {Kunz}, M. and {Kurki-Suonio}, H. and {Lagache}, G. and {Lamarre}, J.-M. and {Lasenby}, A. and {Lattanzi}, M. and {Lawrence}, C.~R. and {Le Jeune}, M. and {Lemos}, P. and {Lesgourgues}, J. and {Levrier}, F. and {Lewis}, A. and {Liguori}, M. and {Lilje}, P.~B. and {Lilley}, M. and {Lindholm}, V. and {L{\'o}pez-Caniego}, M. and {Lubin}, P.~M. and {Ma}, Y.-Z. and {Mac{\'\i}as-P{\'e}rez}, J.~F. and {Maggio}, G. and {Maino}, D. and {Mandolesi}, N. and {Mangilli}, A. and {Marcos-Caballero}, A. and {Maris}, M. and {Martin}, P.~G. and {Martinelli}, M. and {Mart{\'\i}nez-Gonz{\'a}lez}, E. and {Matarrese}, S. and {Mauri}, N. and {McEwen}, J.~D. and {Meinhold}, P.~R. and {Melchiorri}, A. and {Mennella}, A. and {Migliaccio}, M. and {Millea}, M. and {Mitra}, S. and {Miville-Desch{\^e}nes}, M.-A. and {Molinari}, D. and {Montier}, L. and {Morgante}, G. and {Moss}, A. and {Natoli}, P. and {N{\o}rgaard-Nielsen}, H.~U. and {Pagano}, L. and {Paoletti}, D. and {Partridge}, B. and {Patanchon}, G. and {Peiris}, H.~V. and {Perrotta}, F. and {Pettorino}, V. and {Piacentini}, F. and {Polastri}, L. and {Polenta}, G. and {Puget}, J.-L. and {Rachen}, J.~P. and {Reinecke}, M. and {Remazeilles}, M. and {Renzi}, A. and {Rocha}, G. and {Rosset}, C. and {Roudier}, G. and {Rubi{\~n}o-Mart{\'\i}n}, J.~A. and {Ruiz-Granados}, B. and {Salvati}, L. and {Sandri}, M. and {Savelainen}, M. and {Scott}, D. and {Shellard}, E.~P.~S. and {Sirignano}, C. and {Sirri}, G. and {Spencer}, L.~D. and {Sunyaev}, R. and {Suur-Uski}, A.-S. and {Tauber}, J.~A. and {Tavagnacco}, D. and {Tenti}, M. and {Toffolatti}, L. and {Tomasi}, M. and {Trombetti}, T. and {Valenziano}, L. and {Valiviita}, J. and {Van Tent}, B. and {Vibert}, L. and {Vielva}, P. and {Villa}, F. and {Vittorio}, N. and {Wandelt}, B.~D. and {Wehus}, I.~K. and {White}, M. and {White}, S.~D.~M. and {Zacchei}, A. and {Zonca}, A.},
        title = "{Planck 2018 results. VI. Cosmological parameters}",
      journal = {\aap},
         year = 2020,
        month = sep,
       volume = {641},
          eid = {A6},
        pages = {A6},
          doi = {10.1051/0004-6361/201833910},
archivePrefix = {arXiv},
       eprint = {1807.06209},
 primaryClass = {astro-ph.CO},
       adsurl = {https://ui.adsabs.harvard.edu/abs/2020A&A...641A...6P}
}

@ARTICLE{2024PhRvD.109l3530S,
       author = {{Shi}, Yuan and {Zhang}, Pengjie and {Sun}, Zeyang and {Wang}, Yihe},
        title = "{Accurate kappa reconstruction algorithm for masked shear catalog}",
      journal = {\prd},
         year = 2024,
        month = jun,
       volume = {109},
       number = {12},
          eid = {123530},
        pages = {123530},
          doi = {10.1103/PhysRevD.109.123530},
archivePrefix = {arXiv},
       eprint = {2311.00316},
 primaryClass = {astro-ph.CO},
       adsurl = {https://ui.adsabs.harvard.edu/abs/2024PhRvD.109l3530S}
}

@article{van2011numpy,
  title={The NumPy array: a structure for efficient numerical computation},
  author={Van Der Walt, Stefan and Colbert, S Chris and Varoquaux, Gael},
  journal={\CompSE},
  volume={13},
  number={2},
  pages={22--30},
  year={2011},
  doi ={10.1109/MCSE.2011.37},
  publisher={IEEE}
}

@article{harris2020array,
  title={Array programming with NumPy},
  author={Harris, Charles R and Millman, K Jarrod and Van Der Walt, St{\'e}fan J and Gommers, Ralf and Virtanen, Pauli and Cournapeau, David and Wieser, Eric and Taylor, Julian and Berg, Sebastian and Smith, Nathaniel J and others},
  journal={Nature},
  volume={585},
  number={7825},
  pages={357--362},
  doi = {10.1038/s41586-020-2649-2},
  archivePrefix = {arXiv},
  eprint = {2006.10256},
  year={2020},
  publisher={Nature Publishing Group UK London}
}

@article{virtanen2020scipy,
  title={SciPy 1.0: fundamental algorithms for scientific computing in Python},
  author={Virtanen, Pauli and Gommers, Ralf and Oliphant, Travis E and Haberland, Matt and Reddy, Tyler and Cournapeau, David and Burovski, Evgeni and Peterson, Pearu and Weckesser, Warren and Bright, Jonathan and others},
  journal={\natmethod},
  volume={17},
  number={3},
  pages={261--272},
  year={2020},
  doi = {10.1038/s41592-019-0686-2},
  publisher={Nature Publishing Group}
}

@article{robitaille2013astropy,
  title={Astropy: A community Python package for astronomy},
  author={Astropy Collaboration, Robitaille, Thomas P and Tollerud, Erik J and Greenfield, Perry and Droettboom, Michael and Bray, Erik and Aldcroft, Tom and Davis, Matt and Ginsburg, Adam and Price-Whelan, Adrian M and Kerzendorf, Wolfgang E and others},
  journal={\aap},
  volume={558},
  pages={A33},
  year={2013},
  doi={10.1051/0004-6361/201322068},
  publisher={EDP Sciences}
}

@article{price2018astropy,
  title={The astropy project: building an open-science project and status of the v2. 0 core package},
  author={Astropy Collaboration, Price-Whelan, Adrian M and Sip{\H{o}}cz, BM and G{\"u}nther, HM and Lim, PL and Crawford, SM and Conseil, S and Shupe, DL and Craig, MW and Dencheva, N and Ginsburg, A and others},
  journal={\aj},
  volume={156},
  number={3},
  pages={123},
  year={2018},
  doi = {10.3847/1538-3881/aabc4f},
  publisher={IOP Publishing}
}

@article{price2022astropy,
  title={The Astropy Project: sustaining and growing a community-oriented open-source project and the latest major release (v5. 0) of the core package},
  author={Astropy Collaboration, Price-Whelan, Adrian M and Lim, Pey Lian and Earl, Nicholas and Starkman, Nathaniel and Bradley, Larry and Shupe, David L and Patil, Aarya A and Corrales, Lia and Brasseur, CE and N{\"o}the, Maximilian and others},
  journal={\aj},
  volume={935},
  number={2},
  pages={167},
  year={2022},
  doi = {10.3847/1538-4357/ac7c74},
  publisher={IOP Publishing}
}

@article{dalcin2021mpi4py,
  title={mpi4py: Status update after 12 years of development},
  author={Dalcin, Lisandro and Fang, Yao-Lung L},
  journal={\CompSE},
  volume={23},
  number={4},
  pages={47--54},
  year={2021},
  doi={10.1109/MCSE.2021.3083216},
  publisher={IEEE}
}

@article{hunter2007matplotlib,
  title={Matplotlib: A 2D graphics environment},
  author={Hunter, John D},
  journal={\CompSE},
  volume={9},
  number={03},
  pages={90--95},
  year={2007},
  doi = {10.1109/MCSE.2007.55},
  publisher={IEEE Computer Society}
}

@inproceedings{gabriel2004open,
  title={Open MPI: Goals, concept, and design of a next generation MPI implementation},
  author={Gabriel, Edgar and Fagg, Graham E and Bosilca, George and Angskun, Thara and Dongarra, Jack J and Squyres, Jeffrey M and Sahay, Vishal and Kambadur, Prabhanjan and Barrett, Brian and Lumsdaine, Andrew and others},
  booktitle={Recent Advances in Parallel Virtual Machine and Message Passing Interface: 11th European PVM/MPI Users' Group Meeting Budapest, Hungary, September 19--22, 2004. Proceedings 11},
  pages={97--104},
  year={2004},
  url={https://doi.org/10.1007/978-3-540-30218-6_19},
  publisher={Springer}
}

@Book{collette_python_hdf5_2014,
    year = {2013},
    publisher = {O\'Reilly},
    title = {Python and HDF5},
    author = {Andrew Collette}
}

@ARTICLE{2005ApJ...622..759G,
   author = {{G{\'o}rski}, K.~M. and {Hivon}, E. and {Banday}, A.~J. and 
	{Wandelt}, B.~D. and {Hansen}, F.~K. and {Reinecke}, M. and 
	{Bartelmann}, M.},
    title = "{HEALPix: A Framework for High-Resolution Discretization and Fast Analysis of Data Distributed on the Sphere}",
  journal = {\apj},
   eprint = {arXiv:astro-ph/0409513},
     year = 2005,
    month = apr,
   volume = 622,
    pages = {759-771},
      doi = {10.1086/427976},
      url = {https://doi.org/10.1086/427976},
   adsurl = {http://adsabs.harvard.edu/abs/2005ApJ...622..759G}
}

@article{Zonca2019,
  doi = {10.21105/joss.01298},
  url = {https://doi.org/10.21105/joss.01298},
  year = {2019},
  month = mar,
  publisher = {The Open Journal},
  volume = {4},
  number = {35},
  pages = {1298},
  author = {Andrea Zonca and Leo Singer and Daniel Lenz and Martin Reinecke and Cyrille Rosset and Eric Hivon and Krzysztof Gorski},
  title = {healpy: equal area pixelization and spherical harmonics transforms for data on the sphere in Python},
  journal = {\joss}
}

@ARTICLE{2019arXiv191201703P,
       author = {{Paszke}, Adam and {Gross}, Sam and {Massa}, Francisco and {Lerer}, Adam and {Bradbury}, James and {Chanan}, Gregory and {Killeen}, Trevor and {Lin}, Zeming and {Gimelshein}, Natalia and {Antiga}, Luca and {Desmaison}, Alban and {K{\"o}pf}, Andreas and {Yang}, Edward and {DeVito}, Zach and {Raison}, Martin and {Tejani}, Alykhan and {Chilamkurthy}, Sasank and {Steiner}, Benoit and {Fang}, Lu and {Bai}, Junjie and {Chintala}, Soumith},
        title = "{PyTorch: An Imperative Style, High-Performance Deep Learning Library}",
      journal = {arXiv e-prints},
         year = 2019,
        month = dec,
          eid = {arXiv:1912.01703},
        pages = {arXiv:1912.01703},
          doi = {10.48550/arXiv.1912.01703},
archivePrefix = {arXiv},
       eprint = {1912.01703},
 primaryClass = {cs.LG},
       adsurl = {https://ui.adsabs.harvard.edu/abs/2019arXiv191201703P}
}

@article{corner,
  doi = {10.21105/joss.00024},
  url = {https://doi.org/10.21105/joss.00024},
  year  = {2016},
  month = {jun},
  publisher = {The Open Journal},
  volume = {1},
  number = {2},
  pages = {24},
  author = {Daniel Foreman-Mackey},
  title = {corner.py: Scatterplot matrices in Python},
  journal = {\joss}
}

@ARTICLE{2025arXiv250704618C,
       author = {{CSST Collaboration} and {Gong}, Yan and {Miao}, Haitao and {Zhan}, Hu and {Li}, Zhao-Yu and {Shangguan}, Jinyi and {Li}, Haining and {Liu}, Chao and {Chen}, Xuefei and {Yuan}, Haibo and {Zhou}, Jilin and {Liu}, Hui-Gen and {Yu}, Cong and {Ji}, Jianghui and {Qi}, Zhaoxiang and {Liu}, Jiacheng and {Dai}, Zigao and {Wang}, Xiaofeng and {Zheng}, Zhenya and {Hao}, Lei and {Dou}, Jiangpei and {Ao}, Yiping and {Lin}, Zhenhui and {Zhang}, Kun and {Wang}, Wei and {Sun}, Guotong and {Li}, Ran and {Li}, Guoliang and {Xu}, Youhua and {Li}, Xinfeng and {Li}, Shengyang and {Wu}, Peng and {Zhang}, Jiuxing and {Wang}, Bo and {Bai}, Jinming and {Cai}, Yi-Fu and {Cai}, Zheng and {Cao}, Jie and {Chan}, Kwan Chuen and {Chang}, Jin and {Chen}, Xiaodian and {Chen}, Xuelei and {Chen}, Yuqin and {Chen}, Yun and {Cui}, Wei and {Dong}, Subo and {Du}, Pu and {Duan}, Wenying and {Fan}, Junhui and {Fan}, LuLu and {Fan}, Zhou and {Fan}, Zuhui and {Fang}, Taotao and {Fu}, Jianning and {Fu}, Liping and {Fu}, Zhensen and {Gao}, Jian and {Gu}, Shenghong and {Gu}, Yidong and {Guo}, Qi and {Han}, Zhanwen and {Hu}, Bin and {Huang}, Zhiqi and {Ho}, Luis C. and {Jiang}, Linhua and {Jiang}, Ning and {Jing}, Yipeng and {Kang}, Xi and {Kong}, Xu and {Li}, Cheng and {Li}, Chengyuan and {Li}, Di and {Li}, Jing and {Li}, Nan and {Li}, Yang A. and {Liao}, Shilong and {Lin}, Weipeng and {Liu}, Fengshan and {Liu}, Jifeng and {Liu}, Xiangkun and {Liu}, Zhuokai and {Mao}, Ruiqing and {Mao}, Shude and {Meng}, Xianmin and {Pang}, Xiaoying and {Peng}, Xiyan and {Peng}, Yingjie and {Shan}, Huanyuan and {Shen}, Juntai and {Shen}, Shiyin and {Shen}, Zhiqiang and {Shi}, Sheng-Cai and {Shi}, Yong and {Tan}, Siyuan and {Tian}, Hao and {Wang}, Jianmin and {Wang}, Jun-Xian and {Wang}, Xin and {Wang}, Yuting and {Wu}, Hong and {Wu}, Jingwen and {Wu}, Xuebing and {Xu}, Chun and {Xue}, Xiang-Xiang and {Xue}, Yongquan and {Yang}, Ji and {Yang}, Xiaohu and {Yao}, Qijun and {Yuan}, Fangting and {Yuan}, Zhen and {Zhang}, Jun and {Zhang}, Pengjie and {Zhang}, Tianmeng and {Zhang}, Wei and {Zhang}, Xin and {Zhao}, Gang and {Zhao}, Gongbo and {Zhong}, Hongen and {Zhong}, Jing and {Zhou}, Liyong and {Zhu}, Wei and {Zu}, Ying},
        title = "{Introduction to the Chinese Space Station Survey Telescope (CSST)}",
      journal = {arXiv e-prints},
         year = 2025,
        month = jul,
          eid = {arXiv:2507.04618},
        pages = {arXiv:2507.04618},
          doi = {10.48550/arXiv.2507.04618},
archivePrefix = {arXiv},
       eprint = {2507.04618},
 primaryClass = {astro-ph.IM},
       adsurl = {https://ui.adsabs.harvard.edu/abs/2025arXiv250704618C}
}

@ARTICLE{2025SCPMA..6880402G,
       author = {{Gong}, Yan and {Miao}, Haitao and {Zhou}, Xingchen and {Xiong}, Qi and {Song}, Yingxiao and {Jiang}, Yuer and {Wang}, Minglin and {Yan}, Junhui and {Wu}, Beichen and {Deng}, Furen and {Chen}, Xuelei and {Fan}, Zuhui and {Jing}, Yipeng and {Yang}, Xiaohu and {Zhan}, Hu},
        title = "{Future cosmology: New physics and opportunity from the China Space Station Telescope (CSST)}",
      journal = {\scpma},
         year = 2025,
        month = aug,
       volume = {68},
       number = {8},
          eid = {280402},
        pages = {280402},
          doi = {10.1007/s11433-025-2646-2},
archivePrefix = {arXiv},
       eprint = {2501.15023},
 primaryClass = {astro-ph.CO},
       adsurl = {https://ui.adsabs.harvard.edu/abs/2025SCPMA..6880402G}
}

@ARTICLE{2025A&A...697A...1E,
       author = {{Euclid Collaboration} and {Mellier}, Y. and {Abdurro'uf} and {Acevedo Barroso}, J.~A. and {Ach{\'u}carro}, A. and {Adamek}, J. and {Adam}, R. and {Addison}, G.~E. and {Aghanim}, N. and {Aguena}, M. and {Ajani}, V. and {Akrami}, Y. and {Al-Bahlawan}, A. and {Alavi}, A. and {Albuquerque}, I.~S. and {Alestas}, G. and {Alguero}, G. and {Allaoui}, A. and {Allen}, S.~W. and {Allevato}, V. and {Alonso-Tetilla}, A.~V. and {Altieri}, B. and {Alvarez-Candal}, A. and {Alvi}, S. and {Amara}, A. and {Amendola}, L. and {Amiaux}, J. and {Andika}, I.~T. and {Andreon}, S. and {Andrews}, A. and {Angora}, G. and {Angulo}, R.~E. and {Annibali}, F. and {Anselmi}, A. and {Anselmi}, S. and {Arcari}, S. and {Archidiacono}, M. and {Aric{\`o}}, G. and {Arnaud}, M. and {Arnouts}, S. and {Asgari}, M. and {Asorey}, J. and {Atayde}, L. and {Atek}, H. and {Atrio-Barandela}, F. and {Aubert}, M. and {Aubourg}, E. and {Auphan}, T. and {Auricchio}, N. and {Aussel}, B. and {Aussel}, H. and {Avelino}, P.~P. and {Avgoustidis}, A. and {Avila}, S. and {Awan}, S. and {Azzollini}, R. and {Baccigalupi}, C. and {Bachelet}, E. and {Bacon}, D. and {Baes}, M. and {Bagley}, M.~B. and {Bahr-Kalus}, B. and {Balaguera-Antolinez}, A. and {Balbinot}, E. and {Balcells}, M. and {Baldi}, M. and {Baldry}, I. and {Balestra}, A. and {Ballardini}, M. and {Ballester}, O. and {Balogh}, M. and {Ba{\~n}ados}, E. and {Barbier}, R. and {Bardelli}, S. and {Baron}, M. and {Barreiro}, T. and {Barrena}, R. and {Barriere}, J.-C. and {Barros}, B.~J. and {Barthelemy}, A. and {Bartolo}, N. and {Basset}, A. and {Battaglia}, P. and {Battisti}, A.~J. and {Baugh}, C.~M. and {Baumont}, L. and {Bazzanini}, L. and {Beaulieu}, J.-P. and {Beckmann}, V. and {Belikov}, A.~N. and {Bel}, J. and {Bellagamba}, F. and {Bella}, M. and {Bellini}, E. and {Benabed}, K. and {Bender}, R. and {Benevento}, G. and {Bennett}, C.~L. and {Benson}, K. and {Bergamini}, P. and {Bermejo-Climent}, J.~R. and {Bernardeau}, F. and {Bertacca}, D. and {Berthe}, M. and {Berthier}, J. and {Bethermin}, M. and {Beutler}, F. and {Bevillon}, C. and {Bhargava}, S. and {Bhatawdekar}, R. and {Bianchi}, D. and {Bisigello}, L. and {Biviano}, A. and {Blake}, R.~P. and {Blanchard}, A. and {Blazek}, J. and {Blot}, L. and {Bosco}, A. and {Bodendorf}, C. and {Boenke}, T. and {B{\"o}hringer}, H. and {Boldrini}, P. and {Bolzonella}, M. and {Bonchi}, A. and {Bonici}, M. and {Bonino}, D. and {Bonino}, L. and {Bonvin}, C. and {Bon}, W. and {Booth}, J.~T. and {Borgani}, S. and {Borlaff}, A.~S. and {Borsato}, E. and {Bose}, B. and {Botticella}, M.~T. and {Boucaud}, A. and {Bouche}, F. and {Boucher}, J.~S. and {Boutigny}, D. and {Bouvard}, T. and {Bouwens}, R. and {Bouy}, H. and {Bowler}, R.~A.~A. and {Bozza}, V. and {Bozzo}, E. and {Branchini}, E. and {Brando}, G. and {Brau-Nogue}, S. and {Brekke}, P. and {Bremer}, M.~N. and {Brescia}, M. and {Breton}, M.-A. and {Brinchmann}, J. and {Brinckmann}, T. and {Brockley-Blatt}, C. and {Brodwin}, M. and {Brouard}, L. and {Brown}, M.~L. and {Bruton}, S. and {Bucko}, J. and {Buddelmeijer}, H. and {Buenadicha}, G. and {Buitrago}, F. and {Burger}, P. and {Burigana}, C. and {Busillo}, V. and {Busonero}, D. and {Cabanac}, R. and {Cabayol-Garcia}, L. and {Cagliari}, M.~S. and {Caillat}, A. and {Caillat}, L. and {Calabrese}, M. and {Calabro}, A. and {Calderone}, G. and {Calura}, F. and {Camacho Quevedo}, B. and {Camera}, S. and {Campos}, L. and {Ca{\~n}as-Herrera}, G. and {Candini}, G.~P. and {Cantiello}, M. and {Capobianco}, V. and {Cappellaro}, E. and {Cappelluti}, N. and {Cappi}, A. and {Caputi}, K.~I. and {Cara}, C. and {Carbone}, C. and {Cardone}, V.~F. and {Carella}, E. and {Carlberg}, R.~G. and {Carle}, M. and {Carminati}, L. and {Caro}, F. and {Carrasco}, J.~M. and {Carretero}, J. and {Carrilho}, P. and {Carron Duque}, J. and {Carry}, B.},
        title = "{Euclid: I. Overview of the Euclid mission}",
      journal = {\aap},
         year = 2025,
        month = may,
       volume = {697},
          eid = {A1},
        pages = {A1},
          doi = {10.1051/0004-6361/202450810},
archivePrefix = {arXiv},
       eprint = {2405.13491},
 primaryClass = {astro-ph.CO},
       adsurl = {https://ui.adsabs.harvard.edu/abs/2025A&A...697A...1E}
}

@ARTICLE{2019ApJ...873..111I,
       author = {{Ivezi{\'c}}, {\v{Z}}eljko and {Kahn}, Steven M. and {Tyson}, J. Anthony and {Abel}, Bob and {Acosta}, Emily and {Allsman}, Robyn and {Alonso}, David and {AlSayyad}, Yusra and {Anderson}, Scott F. and {Andrew}, John and {Angel}, James Roger P. and {Angeli}, George Z. and {Ansari}, Reza and {Antilogus}, Pierre and {Araujo}, Constanza and {Armstrong}, Robert and {Arndt}, Kirk T. and {Astier}, Pierre and {Aubourg}, {\'E}ric and {Auza}, Nicole and {Axelrod}, Tim S. and {Bard}, Deborah J. and {Barr}, Jeff D. and {Barrau}, Aurelian and {Bartlett}, James G. and {Bauer}, Amanda E. and {Bauman}, Brian J. and {Baumont}, Sylvain and {Bechtol}, Ellen and {Bechtol}, Keith and {Becker}, Andrew C. and {Becla}, Jacek and {Beldica}, Cristina and {Bellavia}, Steve and {Bianco}, Federica B. and {Biswas}, Rahul and {Blanc}, Guillaume and {Blazek}, Jonathan and {Blandford}, Roger D. and {Bloom}, Josh S. and {Bogart}, Joanne and {Bond}, Tim W. and {Booth}, Michael T. and {Borgland}, Anders W. and {Borne}, Kirk and {Bosch}, James F. and {Boutigny}, Dominique and {Brackett}, Craig A. and {Bradshaw}, Andrew and {Brandt}, William Nielsen and {Brown}, Michael E. and {Bullock}, James S. and {Burchat}, Patricia and {Burke}, David L. and {Cagnoli}, Gianpietro and {Calabrese}, Daniel and {Callahan}, Shawn and {Callen}, Alice L. and {Carlin}, Jeffrey L. and {Carlson}, Erin L. and {Chandrasekharan}, Srinivasan and {Charles-Emerson}, Glenaver and {Chesley}, Steve and {Cheu}, Elliott C. and {Chiang}, Hsin-Fang and {Chiang}, James and {Chirino}, Carol and {Chow}, Derek and {Ciardi}, David R. and {Claver}, Charles F. and {Cohen-Tanugi}, Johann and {Cockrum}, Joseph J. and {Coles}, Rebecca and {Connolly}, Andrew J. and {Cook}, Kem H. and {Cooray}, Asantha and {Covey}, Kevin R. and {Cribbs}, Chris and {Cui}, Wei and {Cutri}, Roc and {Daly}, Philip N. and {Daniel}, Scott F. and {Daruich}, Felipe and {Daubard}, Guillaume and {Daues}, Greg and {Dawson}, William and {Delgado}, Francisco and {Dellapenna}, Alfred and {de Peyster}, Robert and {de Val-Borro}, Miguel and {Digel}, Seth W. and {Doherty}, Peter and {Dubois}, Richard and {Dubois-Felsmann}, Gregory P. and {Durech}, Josef and {Economou}, Frossie and {Eifler}, Tim and {Eracleous}, Michael and {Emmons}, Benjamin L. and {Fausti Neto}, Angelo and {Ferguson}, Henry and {Figueroa}, Enrique and {Fisher-Levine}, Merlin and {Focke}, Warren and {Foss}, Michael D. and {Frank}, James and {Freemon}, Michael D. and {Gangler}, Emmanuel and {Gawiser}, Eric and {Geary}, John C. and {Gee}, Perry and {Geha}, Marla and {Gessner}, Charles J.~B. and {Gibson}, Robert R. and {Gilmore}, D. Kirk and {Glanzman}, Thomas and {Glick}, William and {Goldina}, Tatiana and {Goldstein}, Daniel A. and {Goodenow}, Iain and {Graham}, Melissa L. and {Gressler}, William J. and {Gris}, Philippe and {Guy}, Leanne P. and {Guyonnet}, Augustin and {Haller}, Gunther and {Harris}, Ron and {Hascall}, Patrick A. and {Haupt}, Justine and {Hernandez}, Fabio and {Herrmann}, Sven and {Hileman}, Edward and {Hoblitt}, Joshua and {Hodgson}, John A. and {Hogan}, Craig and {Howard}, James D. and {Huang}, Dajun and {Huffer}, Michael E. and {Ingraham}, Patrick and {Innes}, Walter R. and {Jacoby}, Suzanne H. and {Jain}, Bhuvnesh and {Jammes}, Fabrice and {Jee}, M. James and {Jenness}, Tim and {Jernigan}, Garrett and {Jevremovi{\'c}}, Darko and {Johns}, Kenneth and {Johnson}, Anthony S. and {Johnson}, Margaret W.~G. and {Jones}, R. Lynne and {Juramy-Gilles}, Claire and {Juri{\'c}}, Mario and {Kalirai}, Jason S. and {Kallivayalil}, Nitya J. and {Kalmbach}, Bryce and {Kantor}, Jeffrey P. and {Karst}, Pierre and {Kasliwal}, Mansi M. and {Kelly}, Heather and {Kessler}, Richard and {Kinnison}, Veronica and {Kirkby}, David and {Knox}, Lloyd and {Kotov}, Ivan V. and {Krabbendam}, Victor L. and {Krughoff}, K. Simon and {Kub{\'a}nek}, Petr and {Kuczewski}, John and {Kulkarni}, Shri and {Ku}, John and {Kurita}, Nadine R. and {Lage}, Craig S. and {Lambert}, Ron and {Lange}, Travis and {Langton}, J. Brian and {Le Guillou}, Laurent and {Levine}, Deborah and {Liang}, Ming and {Lim}, Kian-Tat and {Lintott}, Chris J. and {Long}, Kevin E. and {Lopez}, Margaux and {Lotz}, Paul J. and {Lupton}, Robert H. and {Lust}, Nate B. and {MacArthur}, Lauren A. and {Mahabal}, Ashish and {Mandelbaum}, Rachel and {Markiewicz}, Thomas W. and {Marsh}, Darren S. and {Marshall}, Philip J. and {Marshall}, Stuart and {May}, Morgan and {McKercher}, Robert and {McQueen}, Michelle and {Meyers}, Joshua and {Migliore}, Myriam and {Miller}, Michelle and {Mills}, David J.},
        title = "{LSST: From Science Drivers to Reference Design and Anticipated Data Products}",
      journal = {\apj},
         year = 2019,
        month = mar,
       volume = {873},
       number = {2},
          eid = {111},
        pages = {111},
          doi = {10.3847/1538-4357/ab042c},
archivePrefix = {arXiv},
       eprint = {0805.2366},
 primaryClass = {astro-ph},
       adsurl = {https://ui.adsabs.harvard.edu/abs/2019ApJ...873..111I}
}

@ARTICLE{2015arXiv150303757S,
       author = {{Spergel}, D. and {Gehrels}, N. and {Baltay}, C. and {Bennett}, D. and {Breckinridge}, J. and {Donahue}, M. and {Dressler}, A. and {Gaudi}, B.~S. and {Greene}, T. and {Guyon}, O. and {Hirata}, C. and {Kalirai}, J. and {Kasdin}, N.~J. and {Macintosh}, B. and {Moos}, W. and {Perlmutter}, S. and {Postman}, M. and {Rauscher}, B. and {Rhodes}, J. and {Wang}, Y. and {Weinberg}, D. and {Benford}, D. and {Hudson}, M. and {Jeong}, W. -S. and {Mellier}, Y. and {Traub}, W. and {Yamada}, T. and {Capak}, P. and {Colbert}, J. and {Masters}, D. and {Penny}, M. and {Savransky}, D. and {Stern}, D. and {Zimmerman}, N. and {Barry}, R. and {Bartusek}, L. and {Carpenter}, K. and {Cheng}, E. and {Content}, D. and {Dekens}, F. and {Demers}, R. and {Grady}, K. and {Jackson}, C. and {Kuan}, G. and {Kruk}, J. and {Melton}, M. and {Nemati}, B. and {Parvin}, B. and {Poberezhskiy}, I. and {Peddie}, C. and {Ruffa}, J. and {Wallace}, J.~K. and {Whipple}, A. and {Wollack}, E. and {Zhao}, F.},
        title = "{Wide-Field InfrarRed Survey Telescope-Astrophysics Focused Telescope Assets WFIRST-AFTA 2015 Report}",
      journal = {arXiv e-prints},
         year = 2015,
        month = mar,
          eid = {arXiv:1503.03757},
        pages = {arXiv:1503.03757},
          doi = {10.48550/arXiv.1503.03757},
archivePrefix = {arXiv},
       eprint = {1503.03757},
 primaryClass = {astro-ph.IM},
       adsurl = {https://ui.adsabs.harvard.edu/abs/2015arXiv150303757S}
}

@ARTICLE{2018PASJ...70S...4A,
       author = {{Aihara}, Hiroaki and {Arimoto}, Nobuo and {Armstrong}, Robert and {Arnouts}, St{\'e}phane and {Bahcall}, Neta A. and {Bickerton}, Steven and {Bosch}, James and {Bundy}, Kevin and {Capak}, Peter L. and {Chan}, James H.~H. and {Chiba}, Masashi and {Coupon}, Jean and {Egami}, Eiichi and {Enoki}, Motohiro and {Finet}, Francois and {Fujimori}, Hiroki and {Fujimoto}, Seiji and {Furusawa}, Hisanori and {Furusawa}, Junko and {Goto}, Tomotsugu and {Goulding}, Andy and {Greco}, Johnny P. and {Greene}, Jenny E. and {Gunn}, James E. and {Hamana}, Takashi and {Harikane}, Yuichi and {Hashimoto}, Yasuhiro and {Hattori}, Takashi and {Hayashi}, Masao and {Hayashi}, Yusuke and {He{\l}miniak}, Krzysztof G. and {Higuchi}, Ryo and {Hikage}, Chiaki and {Ho}, Paul T.~P. and {Hsieh}, Bau-Ching and {Huang}, Kuiyun and {Huang}, Song and {Ikeda}, Hiroyuki and {Imanishi}, Masatoshi and {Inoue}, Akio K. and {Iwasawa}, Kazushi and {Iwata}, Ikuru and {Jaelani}, Anton T. and {Jian}, Hung-Yu and {Kamata}, Yukiko and {Karoji}, Hiroshi and {Kashikawa}, Nobunari and {Katayama}, Nobuhiko and {Kawanomoto}, Satoshi and {Kayo}, Issha and {Koda}, Jin and {Koike}, Michitaro and {Kojima}, Takashi and {Komiyama}, Yutaka and {Konno}, Akira and {Koshida}, Shintaro and {Koyama}, Yusei and {Kusakabe}, Haruka and {Leauthaud}, Alexie and {Lee}, Chien-Hsiu and {Lin}, Lihwai and {Lin}, Yen-Ting and {Lupton}, Robert H. and {Mandelbaum}, Rachel and {Matsuoka}, Yoshiki and {Medezinski}, Elinor and {Mineo}, Sogo and {Miyama}, Shoken and {Miyatake}, Hironao and {Miyazaki}, Satoshi and {Momose}, Rieko and {More}, Anupreeta and {More}, Surhud and {Moritani}, Yuki and {Moriya}, Takashi J. and {Morokuma}, Tomoki and {Mukae}, Shiro and {Murata}, Ryoma and {Murayama}, Hitoshi and {Nagao}, Tohru and {Nakata}, Fumiaki and {Niida}, Mana and {Niikura}, Hiroko and {Nishizawa}, Atsushi J. and {Obuchi}, Yoshiyuki and {Oguri}, Masamune and {Oishi}, Yukie and {Okabe}, Nobuhiro and {Okamoto}, Sakurako and {Okura}, Yuki and {Ono}, Yoshiaki and {Onodera}, Masato and {Onoue}, Masafusa and {Osato}, Ken and {Ouchi}, Masami and {Price}, Paul A. and {Pyo}, Tae-Soo and {Sako}, Masao and {Sawicki}, Marcin and {Shibuya}, Takatoshi and {Shimasaku}, Kazuhiro and {Shimono}, Atsushi and {Shirasaki}, Masato and {Silverman}, John D. and {Simet}, Melanie and {Speagle}, Joshua and {Spergel}, David N. and {Strauss}, Michael A. and {Sugahara}, Yuma and {Sugiyama}, Naoshi and {Suto}, Yasushi and {Suyu}, Sherry H. and {Suzuki}, Nao and {Tait}, Philip J. and {Takada}, Masahiro and {Takata}, Tadafumi and {Tamura}, Naoyuki and {Tanaka}, Manobu M. and {Tanaka}, Masaomi and {Tanaka}, Masayuki and {Tanaka}, Yoko and {Terai}, Tsuyoshi and {Terashima}, Yuichi and {Toba}, Yoshiki and {Tominaga}, Nozomu and {Toshikawa}, Jun and {Turner}, Edwin L. and {Uchida}, Tomohisa and {Uchiyama}, Hisakazu and {Umetsu}, Keiichi and {Uraguchi}, Fumihiro and {Urata}, Yuji and {Usuda}, Tomonori and {Utsumi}, Yousuke and {Wang}, Shiang-Yu and {Wang}, Wei-Hao and {Wong}, Kenneth C. and {Yabe}, Kiyoto and {Yamada}, Yoshihiko and {Yamanoi}, Hitomi and {Yasuda}, Naoki and {Yeh}, Sherry and {Yonehara}, Atsunori and {Yuma}, Suraphong},
        title = "{The Hyper Suprime-Cam SSP Survey: Overview and survey design}",
      journal = {\pasj},
         year = 2018,
        month = jan,
       volume = {70},
          eid = {S4},
        pages = {S4},
          doi = {10.1093/pasj/psx066},
archivePrefix = {arXiv},
       eprint = {1704.05858},
 primaryClass = {astro-ph.IM},
       adsurl = {https://ui.adsabs.harvard.edu/abs/2018PASJ...70S...4A}
}

@ARTICLE{2016MNRAS.460.1270D,
       author = {{Dark Energy Survey Collaboration} and {Abbott}, T. and {Abdalla}, F.~B. and {Aleksi{\'c}}, J. and {Allam}, S. and {Amara}, A. and {Bacon}, D. and {Balbinot}, E. and {Banerji}, M. and {Bechtol}, K. and {Benoit-L{\'e}vy}, A. and {Bernstein}, G.~M. and {Bertin}, E. and {Blazek}, J. and {Bonnett}, C. and {Bridle}, S. and {Brooks}, D. and {Brunner}, R.~J. and {Buckley-Geer}, E. and {Burke}, D.~L. and {Caminha}, G.~B. and {Capozzi}, D. and {Carlsen}, J. and {Carnero-Rosell}, A. and {Carollo}, M. and {Carrasco-Kind}, M. and {Carretero}, J. and {Castander}, F.~J. and {Clerkin}, L. and {Collett}, T. and {Conselice}, C. and {Crocce}, M. and {Cunha}, C.~E. and {D'Andrea}, C.~B. and {da Costa}, L.~N. and {Davis}, T.~M. and {Desai}, S. and {Diehl}, H.~T. and {Dietrich}, J.~P. and {Dodelson}, S. and {Doel}, P. and {Drlica-Wagner}, A. and {Estrada}, J. and {Etherington}, J. and {Evrard}, A.~E. and {Fabbri}, J. and {Finley}, D.~A. and {Flaugher}, B. and {Foley}, R.~J. and {Fosalba}, P. and {Frieman}, J. and {Garc{\'\i}a-Bellido}, J. and {Gaztanaga}, E. and {Gerdes}, D.~W. and {Giannantonio}, T. and {Goldstein}, D.~A. and {Gruen}, D. and {Gruendl}, R.~A. and {Guarnieri}, P. and {Gutierrez}, G. and {Hartley}, W. and {Honscheid}, K. and {Jain}, B. and {James}, D.~J. and {Jeltema}, T. and {Jouvel}, S. and {Kessler}, R. and {King}, A. and {Kirk}, D. and {Kron}, R. and {Kuehn}, K. and {Kuropatkin}, N. and {Lahav}, O. and {Li}, T.~S. and {Lima}, M. and {Lin}, H. and {Maia}, M.~A.~G. and {Makler}, M. and {Manera}, M. and {Maraston}, C. and {Marshall}, J.~L. and {Martini}, P. and {McMahon}, R.~G. and {Melchior}, P. and {Merson}, A. and {Miller}, C.~J. and {Miquel}, R. and {Mohr}, J.~J. and {Morice-Atkinson}, X. and {Naidoo}, K. and {Neilsen}, E. and {Nichol}, R.~C. and {Nord}, B. and {Ogando}, R. and {Ostrovski}, F. and {Palmese}, A. and {Papadopoulos}, A. and {Peiris}, H.~V. and {Peoples}, J. and {Percival}, W.~J. and {Plazas}, A.~A. and {Reed}, S.~L. and {Refregier}, A. and {Romer}, A.~K. and {Roodman}, A. and {Ross}, A. and {Rozo}, E. and {Rykoff}, E.~S. and {Sadeh}, I. and {Sako}, M. and {S{\'a}nchez}, C. and {Sanchez}, E. and {Santiago}, B. and {Scarpine}, V. and {Schubnell}, M. and {Sevilla-Noarbe}, I. and {Sheldon}, E. and {Smith}, M. and {Smith}, R.~C. and {Soares-Santos}, M. and {Sobreira}, F. and {Soumagnac}, M. and {Suchyta}, E. and {Sullivan}, M. and {Swanson}, M. and {Tarle}, G. and {Thaler}, J. and {Thomas}, D. and {Thomas}, R.~C. and {Tucker}, D. and {Vieira}, J.~D. and {Vikram}, V. and {Walker}, A.~R. and {Wechsler}, R.~H. and {Weller}, J. and {Wester}, W. and {Whiteway}, L. and {Wilcox}, H. and {Yanny}, B. and {Zhang}, Y. and {Zuntz}, J.},
        title = "{The Dark Energy Survey: more than dark energy - an overview}",
      journal = {\mnras},
         year = 2016,
        month = aug,
       volume = {460},
       number = {2},
        pages = {1270-1299},
          doi = {10.1093/mnras/stw641},
archivePrefix = {arXiv},
       eprint = {1601.00329},
 primaryClass = {astro-ph.CO},
       adsurl = {https://ui.adsabs.harvard.edu/abs/2016MNRAS.460.1270D}
}

@ARTICLE{2023A&A...673A.111Y,
       author = {{Yao}, Ji and {Shan}, Huanyuan and {Zhang}, Pengjie and {Liu}, Xiangkun and {Heymans}, Catherine and {Joachimi}, Benjamin and {Asgari}, Marika and {Bilicki}, Maciej and {Hildebrandt}, Hendrik and {Kuijken}, Konrad and {Tr{\"o}ster}, Tilman and {van den Busch}, Jan Luca and {Wright}, Angus and {Yan}, Ziang},
        title = "{KiDS-1000: Cross-correlation with Planck cosmic microwave background lensing and intrinsic alignment removal with self-calibration}",
      journal = {\aap},
         year = 2023,
        month = may,
       volume = {673},
          eid = {A111},
        pages = {A111},
          doi = {10.1051/0004-6361/202346020},
archivePrefix = {arXiv},
       eprint = {2301.13437},
 primaryClass = {astro-ph.CO},
       adsurl = {https://ui.adsabs.harvard.edu/abs/2023A&A...673A.111Y}
}

@article{HYVARINEN2000411,
title = {Independent component analysis: algorithms and applications},
journal = {Neural Networks},
volume = {13},
number = {4},
pages = {411-430},
year = {2000},
issn = {0893-6080},
doi = {https://doi.org/10.1016/S0893-6080(00)00026-5},
url = {https://www.sciencedirect.com/science/article/pii/S0893608000000265},
author = {A. Hyvärinen and E. Oja}
}


\clearpage
\onecolumn

\begin{appendix}
\section{SBI Results}
\label{sec: appendix}

This Appendix presents the full set of inferred posteriors from 10 independent SBI realizations, together with 6 types of summary statistics (Table~\ref{tab: cosmological constraints}), for a fiducial cosmology with $\Omega_{\rm m}=0.3062$ and $\sigma_8=0.7615$ over $655.4~{\rm deg}^2$.

\begin{figure}[H]
\centering
\includegraphics[width=0.7\textwidth]{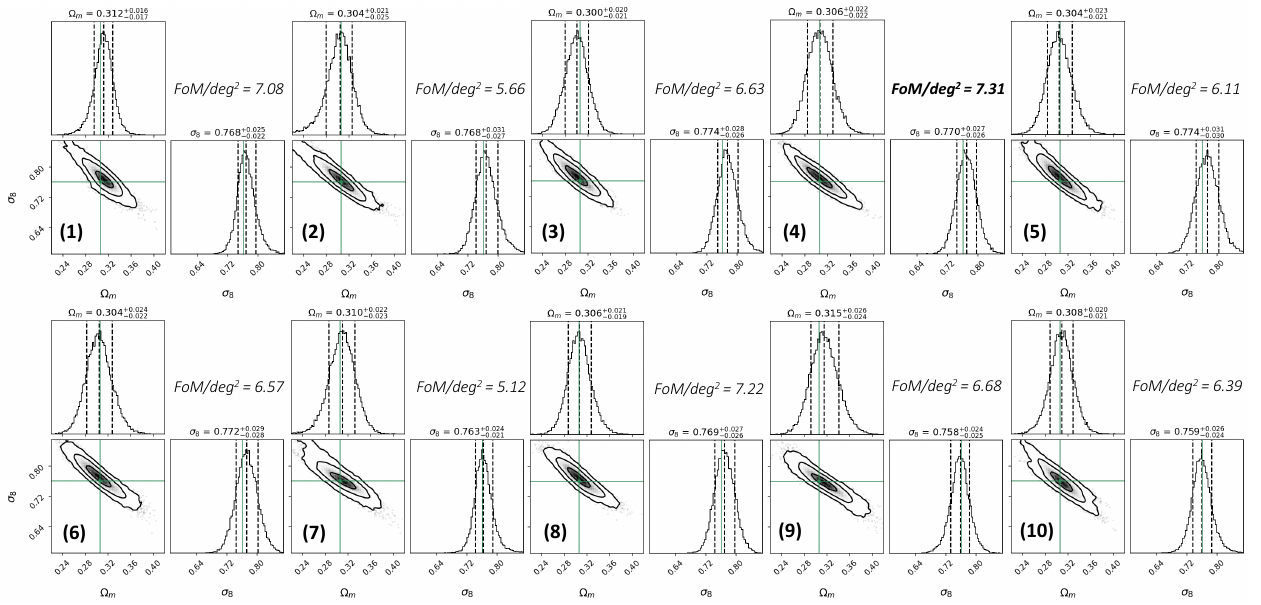}
\caption{\label{fig: gamma-MsmAvg-2PCF-10SBI-runs} 
SBI posteriors based on 2PCF ($\hat{\xi}_{+-}$) of $\gamma^{\rm Avg}$.}
\end{figure}

\begin{figure}[H]
\centering
\includegraphics[width=0.7\textwidth]{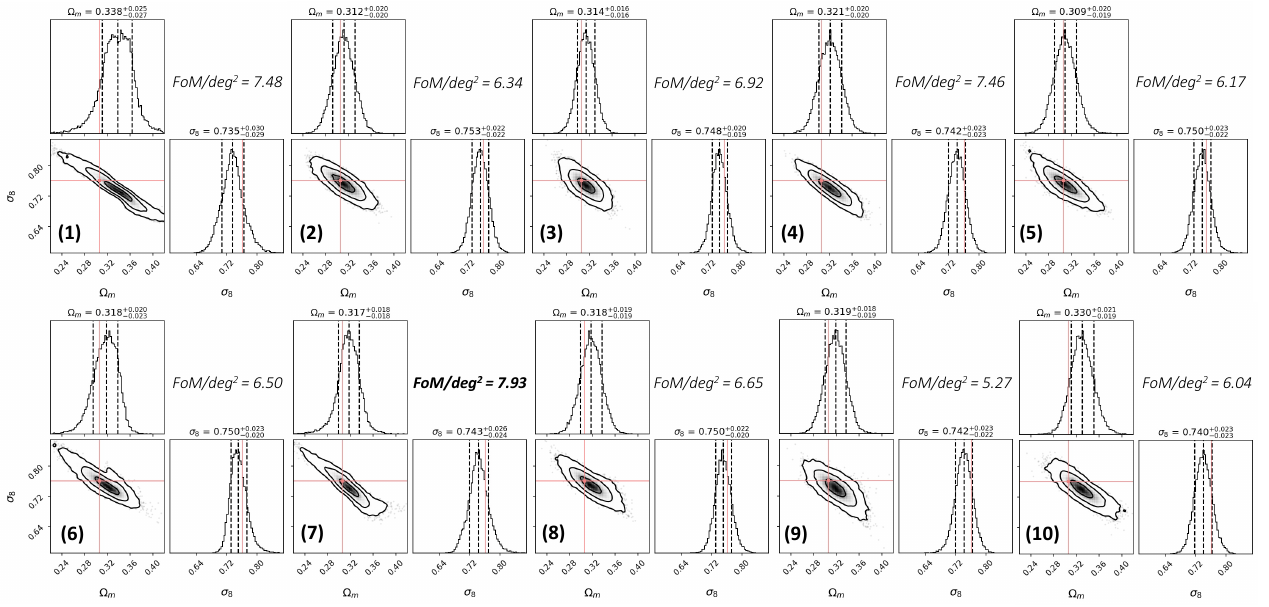}
\caption{\label{fig: gamma-MsmAvg-10SBI-runs} 
SBI posteriors based on ML-derived features ($x_{\texttt{ML}}$) of $\gamma^{\rm Avg}$.}
\end{figure}

\begin{figure}[H]
\centering
\includegraphics[width=0.7\textwidth]{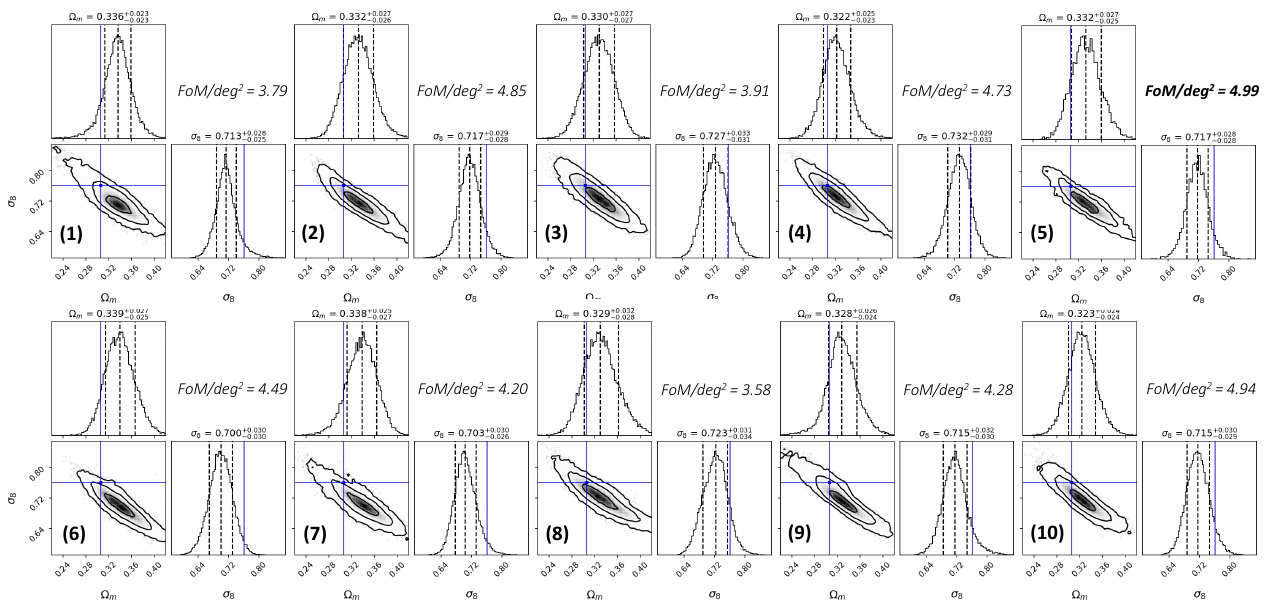}
\caption{\label{fig: kappa-MsmAvg-10SBI-runs} 
SBI posteriors based on ML-derived features ($x_{\texttt{ML}}$) of $\kappa^{\rm Avg}$.}
\end{figure}

\begin{figure}[H]
\centering
\includegraphics[width=0.7\textwidth]{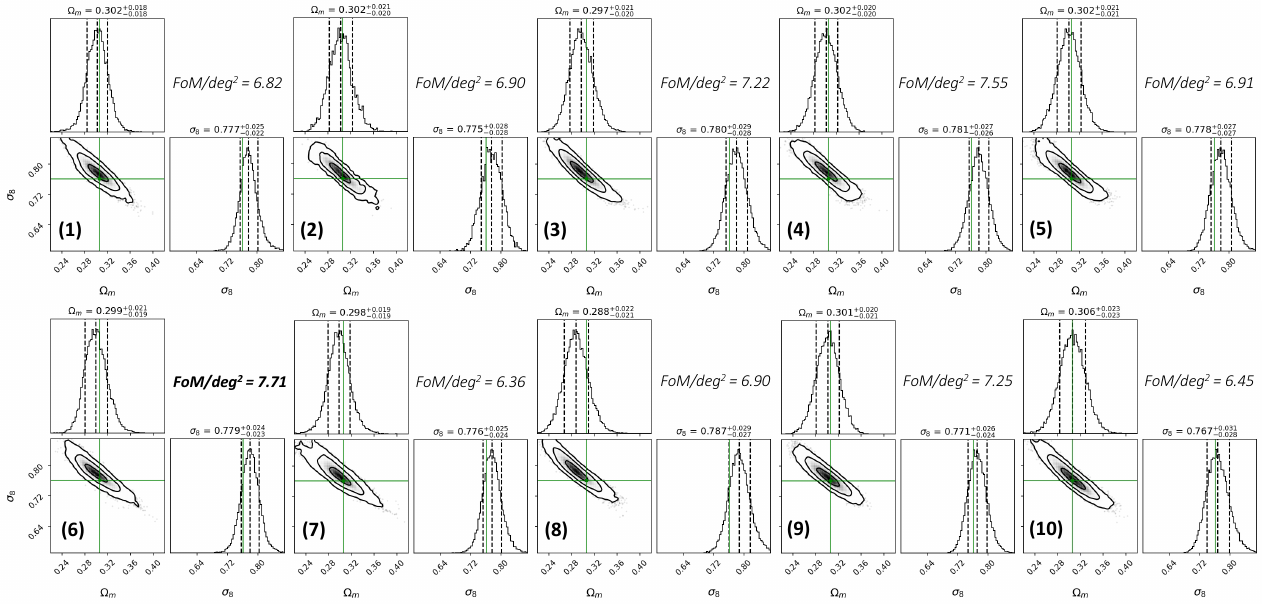}
\caption{\label{fig: gamma-MsmPCA-2PCF-10SBI-runs} 
SBI posteriors based on 2PCF ($\hat{\xi}_{+-}$) of $\gamma^{\rm PCA}$.}
\end{figure}

\begin{figure}[H]
\centering
\includegraphics[width=0.7\textwidth]{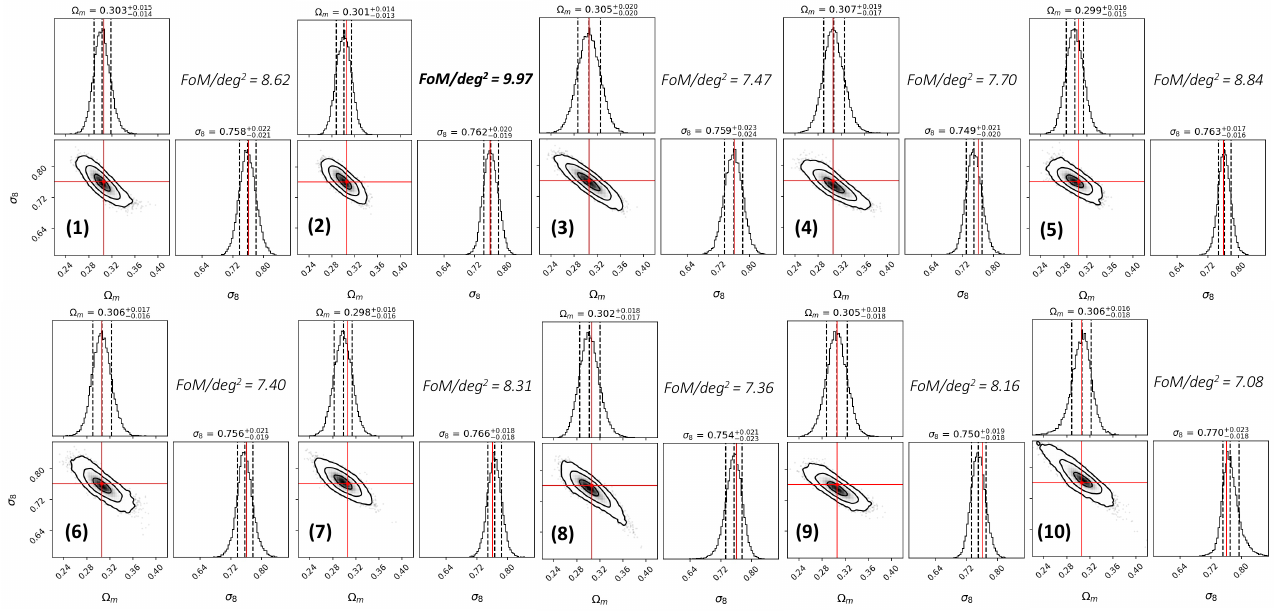}
\caption{\label{fig: gamma-MsmPCA-10SBI-runs} 
SBI posteriors based on ML-derived features ($x_{\texttt{ML}}$) of $\gamma^{\rm PCA}$.}
\end{figure}

\begin{figure}[H]
\centering
\includegraphics[width=0.7\textwidth]{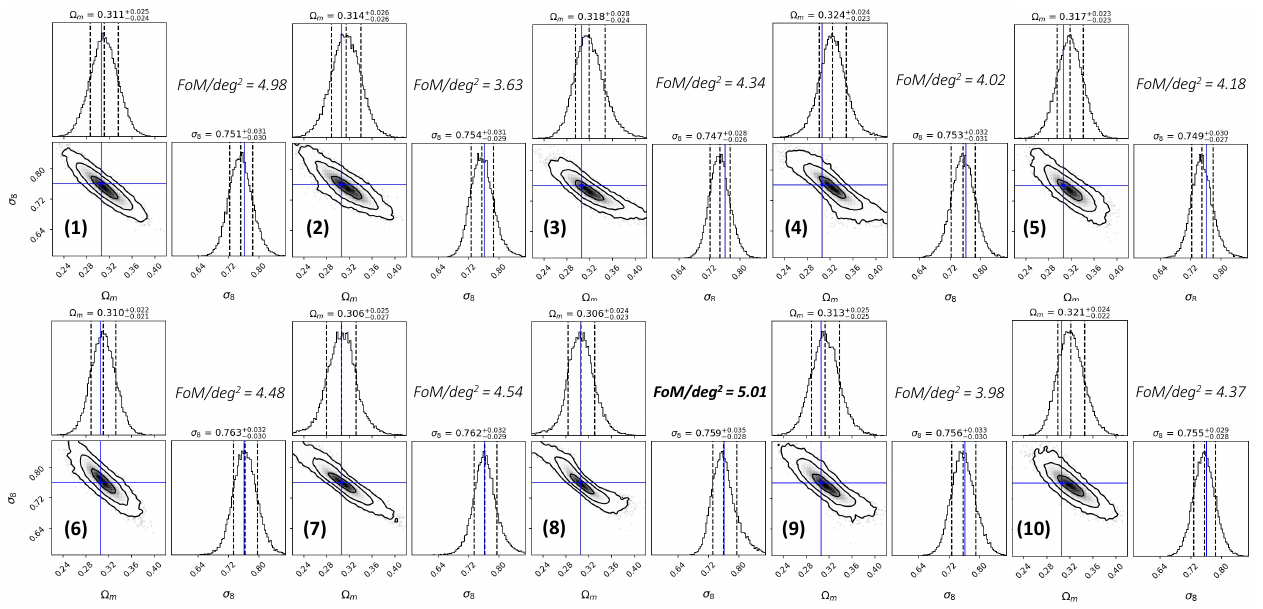}
\caption{\label{fig: kappa-MsmPCA-10SBI-runs} 
SBI posteriors based on ML-derived features ($x_{\texttt{ML}}$) of $\kappa^{\rm PCA}$.}
\end{figure}

\end{appendix}


\end{document}